\documentclass[twocolumn,showpacs,amsmath,amssymb,pra,longbibliography]{revtex4-1}

\usepackage[pdftex]{graphicx} 
\usepackage{dcolumn}  
\usepackage{bm}       
\usepackage[usenames,dvipsnames]{color}
\definecolor{URLCOL}{rgb}{0,0.52,0.83} 
\definecolor{LINKCOL}{rgb}{0.05,0.5,0} 
\definecolor{orange}{rgb}{0.6,0.3,0} 
\definecolor{CITECOL}{rgb}{0.25,0,0.48} 
\usepackage{epstopdf}

\usepackage[pdftex,bookmarks,breaklinks,bookmarksopen,bookmarksnumbered,colorlinks,linkcolor=LINKCOL,linktocpage,citecolor=CITECOL,urlcolor=URLCOL,pdfpagemode=UseOutline,pdftex]{hyperref}

\usepackage{booktabs}
\usepackage{comment}
\usepackage{natbib}

\definecolor{TITLECOL}{rgb}{0.1,0.2,0.7} 
\definecolor{SECOL}{rgb}{0.1,0.2,0.7} 
\definecolor{CONTENTSCOL}{rgb}{0.1,0.2,0.7} 
\definecolor{SSECOL}{rgb}{0.25,0,0.48} 
\definecolor{SSSECOL}{rgb}{0.2,0.08,0.53} 
\definecolor{FINCOL}{rgb}{0.01,0.3,0.07} 

\def\coltableofcontents{ 
{
\definecolor{SECOL}{rgb}{0.25,0,0.48} 
\definecolor{SSECOL}{rgb}{0.2,0.08,0.53} 
\tableofcontents
}
}

\def\erf{\text{erf }}

\def\shd#1{\vskip0.35cm {\bf \color{SSSECOL}#1:}} 

\def\coloredtitle#1{\title{\textcolor{TITLECOL}{#1}}} 
\def\coloredauthor#1{\author{\textcolor{CITECOL}{#1}}} 

\definecolor{URLCOL}{rgb}{0,0.17,0.43} 
\definecolor{LINKCOL}{rgb}{0.05,0.4,0} 
\definecolor{CITECOL}{rgb}{0.35,0,0.48} 

\def\sss{\scriptscriptstyle\rm}

\def\bea{\begin{eqnarray}}
\def\eea{\end{eqnarray}}
\def\ben{\begin{equation}}
\def\een{\end{equation}}
\def\benu{\begin{enumerate}}
\def\enu{\end{enumerate}}

\def\bei{\begin{itemize}}
\def\eei{\end{itemize}}
\def\beit{\begin{itemize}}
\def\eit{\end{itemize}}
\def\benu{\begin{enumerate}}
\def\enu{\end{enumerate}}

\def\heavi{\theta}
\def\hatT{{\hat T}}
\def\hatV{{\hat V}}
\def\hatH{{\hat H}}
\def\hatW{{\hat W}}
\def\br{{\bf r}}

\def\bR{{\bf R}}
\def\bu{{\bf u}}
\def\bff{{\bf f}}

\def\half{\frac{1}{2}}

\def\x{_{\sss X}}

\def\c{_{\sss C}}
\def\s{_{\sss S}}

\def\xc{_{\sss XC}}

\def\Hxc{_{\sss HXC}}
\def\H{_{\sss H}}

\def\KSLDA{^{\rm KS\text{-}LDA}}

\def\NL{^{\rm NLR}}
\def\Hartree{^{\rm Hartree}}
\def\HYB{^{\rm hyb}}

\def\KSNL{^{\rm KS\text{-}NLR}}

\def\homo{_{\rm HOMO}}

\def\NLunif{^{\rm NL\text{-}unif}}
\def\PCunif{^{\rm PC\text{-}unif}}
\def\SCEunif{^{\rm SCE\text{-}unif}}
\def\up{_\uparrow}
\def\dn{_\downarrow}

\def\intr{\int d^3r\,}
\def\intusph{\int_{4\pi} \!\! d\Omega_{\hat u}\,}
\def\intsphu{\dfrac{1}{4\pi} \intusph}

\def\intu{\int d^3u\,}
\def\intrp{\int d^3r'\,}
\def\intOrp{\int_{\Omega(\br)}\!\!\!\!\! d^3r'\,}
\def\intrOrp{\intr \! \intOrp}
\def\intdOr{\oint_{\partial \Omega}\!\! d{\bf A}(\br) \cdot}
\def\intOrrp{\int_{\Omega(\br)}\!\!\!\!\! d^3r''\,}
\def\intSrp{\int_{\partial \Omega(\br)}\!\!\!\!\!\!\! d^2r'\,}
\def\intrrp{\int d^3r \int d^3r'\,}

\def\Iomega{\widetilde{ \Omega}}

\def\n{n}

\def\PC{^\text{PC}} 
\def\PCGGA{^\text{PC-GGA}} 
\def\sce{_\text{SCE}}

\def\SCE{^\text{SCE}}

\def\Tabref#1{Table~\ref{#1}}

\def\Eqref#1{Eq.~\eqref{#1}}

\def\Secref#1{Section~\ref{#1}}
\def\Appref#1{Appendix~\ref{#1}}
\def\Figref#1{Fig.~\ref{#1}}
\def\Ref#1{Ref.~\cite{#1}}
\def\Refs#1{Refs.~\cite{#1}}

\def\sec#1{\section{\textcolor{SECOL}{#1}}}
\def\ssec#1{\subsection{\textcolor{SSECOL}{#1}}}

\begin{document}

\coloredtitle{
How to make electrons avoid each other:
a nonlocal radius for strong correlation
}
\coloredauthor{Lucas O.\ Wagner}
\coloredauthor{Paola Gori-Giorgi}
\affiliation
{Department of Theoretical Chemistry and Amsterdam Center 
for Multiscale Modeling, FEW, Vrije Universiteit, De Boelelaan 1083, 1081HV Amsterdam, 
The Netherlands}

\date{\today}

\begin{abstract}
We present here a model of the exchange-correlation hole for strongly correlated systems
using a simple nonlocal generalization of the Wigner--Seitz radius.  The model
behaves similarly to the strictly correlated electron approach, which gives the infinitely
correlated limit of density functional theory.  Unlike the strictly correlated method, however,
the energies and potentials of this model can be presently calculated for arbitrary
geometries in three dimensions.  We discuss how to evaluate the energies and potentials of 
the nonlocal model, and
provide results for many systems where it is also possible to compare to the strictly
correlated electron treatment.
\end{abstract}


\maketitle
\coltableofcontents


\sec{Introduction}

Treating strong electronic correlation at an affordable computational cost is an important missing building block for a truly predictive computational material science, chemistry, and biochemistry \cite{CohMorYan-CR-12,Bec-JCP-14}.
Though each strongly correlated system seems to require a new model and method, there are some powerful
tools for probing these systems. Smart wave function methods such as the density matrix renormalization group \cite{White:1992,White:1993a} 
yield impressive results for lattice models \cite{Schollwoeck:2005}
and enjoy some success in describing real continuum systems \cite{White:1999,Chan:2008,CS11},
but they are still computationally expensive, and thus limited as far as system size is concerned.
On the other hand, the much cheaper Kohn--Sham density functional theory (KS-DFT) \cite{KS65} is often
considered a method suited only for weakly correlated systems. 
The very first approximation in KS-DFT, the local density approximation (LDA) \cite{KS65},
describes well the physics and energies of weakly correlated electrons,
but can fail spectacularly for strongly correlated systems,
where the wavefunction is radically different than that of
the KS non-interacting reference system.
(In some communities, failure of KS-LDA is taken as the
{\em definition} of strong correlation.) 
All standard KS-DFT approximations build upon LDA in one way or another \cite{PBE96,B88,Bb93,HSE06},
achieving greater accuracy in weakly correlated systems.
In systems where KS-LDA is qualitatively wrong, however, fine-tuned improvements seem to have little hope
of succeeding. The situation appears rather dire then to describe strong correlation with KS-DFT.

Despite this bleak situation, recent work has uncovered a functional which is correct in the
strongly correlated limit -- the strictly correlated electron (SCE) functional
\cite{S99,SGS07,GorVigSei-JCTC-09}, 
defined as the minimum possible expectation value of the electron-electron 
repulsion in a given electron density.
The SCE functional depends in a highly nonlocal way
on the density \cite{MalGor-PRL-12,MalMirCreReiGor-PRB-13}.
This nonlocality is a new rung in the
Jacob's ladder strategy  \cite{PerSmi-INC-01,PRTS04} for constructing more accurate approximate 
functionals, by using information of increasing complexity.
Traditionally, the rungs include:
local information (the value of the density at each point in space, yielding e.g., LDA, LSDA \cite{KS65,PW92}), 
semilocal information (density gradients, giving GGA's \cite{PBE96,B88,LYP88}), 
the local KS kinetic energy density (metaGGA's \cite{TPSS03}), 
the occupied KS orbitals (hybrids \cite{BEP97,Bb93}, self-interaction corrections \cite{PZ81,PRP14}, etc.), 
up to the KS virtuals (double hybrids \cite{Gri-JCP-06}, 
random-phase approximation in different flavours \cite{PRRS12}, etc.).
As we will see, the nonlocal rung utilizes information about certain {\em integrals}
of the density; there are other such functionals (e.g.,
the weighted density approximation \cite{AG77,AG78,GJL77,GJL79,DG90}), and we will
introduce another in this work.

When applied in the self-consistent KS-DFT framework \cite{MalGor-PRL-12}, 
the KS-SCE method correctly describes strong correlation
phenomena such as bond dissociation \cite{MalMirGieWagGor-PCCP-14} and charge localization in one-dimensional (1d) and two-dimensional (2d) 
traps with weak confinement 
\cite{MalMirCreReiGor-PRB-13,MenMalGor-PRB-14}, without introducing any artificial symmetry breaking.
There are still challenges to meet before KS-SCE is ready for practical applications, however.
In the three dimensional (3d) case, algorithms to compute the SCE functional are currently
applicable only to spherically symmetric systems, although progress is being made by several
groups exploiting the formal similarity between the SCE problem and optimal transport
(or mass transportation) theory \cite{ButDepGor-PRA-12,Pas-JFA-13,CotFriKlu-CPAM-13,MenLin-PRB-13,CFM14,FriMenPasCotKlu-JCP-13}, a
field of mathematics and economics \cite{Vil-BOOK-03}. 
Another crucial point is that SCE requires suitable corrections 
\cite{MalMirGieWagGor-PCCP-14,MirUmrMorGor-JCP-14} (e.g., local or semilocal) to make it useful 
for chemistry and solid state physics, where both the strong and the weak correlation regimes need to be treated accurately by the same methodology.
Despite these challenges, hybridizing with the SCE functional (or an approximation thereof) holds the promise
of remedying strong correlation failures in present KS-DFT functionals.

In this work, we present an approximation to the SCE functional, 
which is easier to implement for arbitrary geometries and may more readily
accept corrections.  The primary ingredient for this
functional is a nonlocal generalization of the Wigner--Seitz radius.
With this nonlocal Wigner--Seitz radius, 
we build a very natural (and nonlocal) model of
the strong-interaction limit  of the exchange-correlation hole.
We call the resulting functional the nonlocal radius (NLR) functional.
Previous attempts to build approximations for the SCE limit were based on
local or semilocal information \cite{SPK00b,SeiPerKur-PRL-00}.  
These models can be energetically
accurate, but they miss nonlocality in the functional derivative;
nonlocality which is 
necessary to self-consistently build  features such as barriers 
that localize the charge density \cite{MalGor-PRL-12,MalMirCreReiGor-PRB-13,MenMalGor-PRB-14}. 
With the nonlocal Wigner--Seitz radius,
our functional captures, self-consistently, some of the physics of strong correlation.

The paper is organized as follows. Since the exchange-correlation hole \cite{B97b,GorAngSav-CJC-09} 
and the strictly correlated electron approach are both useful in understanding our model, 
we discuss some relevant background on KS-DFT, SCE, and the exchange-correlation hole in \Secref{background}. 
We then 
present the new nonlocal KS-NLR functional and its properties in \Secref{newhole}. 
The KS-NLR method is similar to some other nonlocal approximations, such as
the weighted density approximation \cite{AG77,AG78,GJL77,GJL79,DG90}, 
and may be considered as a (nonlocal) simplification  of the point-charge plus continuum (PC)
model \cite{SPK00b,Ons-JPC-39}, so we discuss these relationships also in \Secref{newhole}.
Finally, we compare self-consistent KS-SCE and KS-NLR calculations for a few systems in \Secref{results},
where it becomes clear that these two methods behave quite similarly.

Though the nonlocal aspect of the KS-NLR functional offers its own set of integration challenges,
evaluating the functional and its functional derivative with respect to the electron density
is straightforward.  To show this, we calculate the nonlocal functional
for real 3d atoms---self-consistently for two electrons, and non-self-consistently
for more---where the KS-SCE functional has also been evaluated \cite{MSG12}.  We also
treat the 3d hydrogen molecule non-self-consistently, 
which has only recently been treated by the exact KS-SCE functional \cite{CFM14,VucWagMirGor-prep-14}.
In comparing the KS-NLR and KS-SCE methods,
we also consider a simplified 1d universe, as in \Ref{WSBW12}, where self-consistent calculations 
can be carried out very rapidly.  There we discover that the nonlocal functional is capable of
dissociating a single bond (such as in H$_2$ or LiH) correctly, as well as localizing charge density
in 1d parabolic traps without symmetry breaking, just like the KS-SCE method \cite{MalMirCreReiGor-PRB-13}.
We therefore consider the KS-NLR approach a presently viable alternative to KS-SCE that opens
many doors in the development of strongly correlated KS-DFT methods. 

\sec{Background}\label{background}

Most investigations into electronic structure begin with the Born--Oppenheimer approximation.
This allows the quantum electronic problem to be solved first, and then any quantum
(or classical) nuclear effects to be added in later.  The electronic Hamiltonian is:
\ben
\hatH = \hatT + \hatW + \hatV,\label{H}
\een
with operators for the kinetic energy, $\hatT$, electron-electron
interaction, $\hatW$, and potential energy, $\hatV$.  
For a system of $N$ electrons, these quantities may be written in the position
representation as (using atomic units):
\ben
\quad
\left.
\begin{array}{rcl}
\hatT &\equiv& \displaystyle -\sum_{i=1}^N \half \nabla_i^2 \\[12pt]
\hatW &\equiv& \displaystyle \sum_{j>i}^N \dfrac{1}{|\br_i - \br_j|} \\[12pt]
\hatV &\equiv& \displaystyle \sum_{i=1}^N v(\br_i)
\end{array}
\quad\quad
\right\}
\label{Hoperators}
\een
where $v(\br)$ is the external potential usually coming from classical nuclei:
\ben
v(\br) = -\sum_{\alpha} \dfrac{Z_\alpha}{|\br - \bR_\alpha|}
\een
where $Z_\alpha$ ($\bR_\alpha$) is the charge (position) of the 
$\alpha$th nucleus.
Minimizing \Eqref{H} over properly antisymmetrized 
wavefunctions yields the ground-state electron wavefunction $\Psi$, which is the key to
many properties of the system.  

Due to theorems by Hohenberg and Kohn \cite{HK64}, we can write the expectation
values of all the operators in \Eqref{Hoperators} as functionals of the electron
density $\n(\br)$.
For $\hatT$ and $\hatW$, 
this is accomplished by the constrained search formalism \cite{L79,L83},
where the internal energy of the system (kinetic plus electron-electron repulsion) is minimized over
wavefunctions $\Psi$ constrained to yield the density $\n(\br)$:
\ben
F[\n] \equiv T[\n] + W[\n] \equiv \min_{\Psi \to \n} \langle \Psi | \hatT + \hatW | \Psi \rangle, \label{F}
\een
where the minimizing $\Psi$ is denoted $\Psi[\n]$. 
For systems with degeneracy, a suitable generalization to mixed states
$\Psi[\n] \to \Gamma[\n]$ is required, with a trace replacing the bra-ket in \Eqref{F}
\cite{V80,L82,UK01,L03,WSBW13}.
The ground-state energy and density are then obtained through a minimization over reasonable densities \cite{L83}
integrating to a certain desired particle number $N$:
\bea
E_v[\n] &\equiv& T[\n] + W[\n] + \intr \n(\br)\, v(\br) \label{Evn}\\
E_v(N) &\equiv& \min_{\n\to N} E_v[\n] \label{EvN}.
\eea

\shd{Kohn--Sham DFT}
The most widely applied DFT is Kohn--Sham DFT (KS-DFT) \cite{KS65},
which uses a set of fictictious non-interacting electrons to capture (in part) the
Fermi statistics 
of the real system.
In Kohn--Sham theory, the energy of \Eqref{Evn} is partitioned as
\ben
E_v[\n] \equiv T\s[\n] + \intr \n(\br)\, v(\br) + E\Hxc[\n],\label{EKS}
\een
where $T\s[\n]$ is the kinetic energy of a set of non-interacting electrons
with density $\n(\br)$,
and $E\Hxc[\n]$ is the Hartree-exchange-correlation energy which
incorporates all effects due to electron interaction.  

The functional derivative of \Eqref{EKS} reveals a gradient-descent
procedure for minimizing $E_v[\n]$ \cite{WSBW13}, and leads to a set of equations which
must be solved self consistently for the electron density $\n(\br)$:
\bea
v(\br) + v\Hxc[\n](\br) &=& v\s(\br) \label{vsupdate}\\[5pt]
\left\{-\half \nabla^2 + v\s(\br) \right\} \phi_{j}(\br) &=& \epsilon_{j}\, \phi_{j}(\br) \label{KSeq}\\ 
2\sum_{j=1}^{N/2} |\phi_j(\br)|^2 &=& \n(\br), \label{KSdens}
\eea
where $v\Hxc[\n](\br)$ is the Hartree-exchange-correlation (HXC) potential,
the functional derivative of $E\Hxc[\n]$:
\ben
v\Hxc[\n](\br) \equiv \dfrac{\delta E\Hxc[\n] }{ \delta \n(\br)},
\een
and, for simplicity, we have considered a spin-unpolarized system (equal numbers of spin-up and 
spin-down electrons:  $N\up = N\dn = N/2$).
In the KS scheme, one solves the above KS equations iteratively, 
until self-consistency is achieved \cite{DG90}.
The ground-state energy may then be computed 
by evaluating $E\Hxc[\n]$ with the converged density and the KS kinetic energy $T\s[\n]$ using the converged
KS orbitals:
\ben
T\s[\n] = - \sum_{j=1}^{N/2} \intr \phi_{j}^*(\br) \nabla^2 \phi_{j}(\br). \label{Ts}
\een
For non-self-consistent densities, $T\s[\n]$ must be evaluated using some other
approach, e.g.\ via inversions \cite{G92,WP93,LB94,ZMP94,LGB96,SGB97,PNW03,WBSB14}.

So far we have only rewritten the original electronic structure problem;
with the exact $E\Hxc[\n]$ functional, KS calculations are even more difficult than
solving the electronic Hamiltonian in \Eqref{H} directly \cite{WSBW13}.
The overwhelming practical success of KS-DFT is that $E\Hxc[\n]$ breaks up into pieces which can be more
easily modeled.  One can split $W[\n]$ and $T[\n]$ into pieces, which can be 
reassembled into $E\Hxc[\n]$:
\ben
\left.
\quad
\begin{array}{rcl}
W[\n] &\equiv&  W\H[\n] + W\xc[\n] \\[8pt]
T[\n] &\equiv&  T\s[\n] + T\c[\n]
\end{array}
\quad
\right\} \label{WTbreakdown}
\een
where $W\H[\n]$ is the Hartree energy of a density $\n(\br)$, often denoted by
$E\H[\n]$ or $U[\n]$, $W\xc[\n]$ is the interaction XC energy,
composed of the exchange energy $E\x[\n] = W\x[\n]$ and the interaction
correlation energy $W\c[\n]$,
and $T\c[\n]$ is the kinetic correlation energy \footnote{%
Sometimes $W\c[\n]$ of \Eqref{WTbreakdown} is called the {\em potential} correlation energy,
or written $U\c[\n]$.  However, the authors consider {\em interaction} correlation energy
to more appropriately describe the correlation energy due to the Coulomb considerations.
It must be clarified, however, that both $T\c[\n]$ and
$W\c[\n]$ are a result of electron interaction:
$T\c[\n]$ is the increase in kinetic energy
due to electrons avoiding each other more, while $W\c[\n]$ is the decrease in Coulomb
interaction energy.
}.
The full correlation energy $E\c[\n]$ contains both kinetic and interaction contributions:
\ben
E\c[\n] = W\c[\n] + T\c[\n],
\een
but no simple explicit expression exists for either (or both) of these terms.
Finally, one obtains $E\Hxc[\n] = E\H[\n] + E\xc[\n]$,
with $E\xc[\n] = E\x[\n] + E\c[\n]$.

\shd{Energies from the XC hole}
All of the interaction terms  in \Eqref{WTbreakdown} can be naturally written as 
Coulomb integrals.  The true $W[\n]$ can be found using the pair density $P(\br,\br')$ of the interacting system:
\bea
&&\quad P(\br,\br') =  N(N-1) \times \label{pair}\\
&&\ \ \sum_{\sigma_1,\ldots,\sigma_N}\int d^3r_3\cdots d^3r_N\, 
    |\Psi(\br \sigma_1,\br' \sigma_2,\br_3 \sigma_3,\ldots, \br_N \sigma_N)|^2,\nonumber
\eea
with the full interaction energy being:
\ben
W[\n] = \half \intrrp \dfrac{P(\br,\br')}{|\br - \br'|}. \label{W}
\een
Unfortunately, $P(\br,\br')$ cannot be varied directly to determine $W[\n]$
(though see \Ref{AL05} for an excellent discussion).   
Instead, the pair density must be modeled in some way,
and in KS-DFT this is done through the density.  
The pair density is thus broken
up into the following terms which are easier to model:
\ben
P(\br,\br') = \n(\br) \big[ \n(\br') + h\xc(\br,\br') \big] \label{pairhole}
\een
where $h\xc(\br,\br')$ is the exchange-correlation hole, which has some 
simple properties summarized below.

Plugging $P(\br,\br')$ \eqref{pairhole} into $W[\n]$ \eqref{W}, 
we can obtain the various interaction energies
of \Eqref{WTbreakdown}.  The Hartree piece is
\ben
W\H[\n] \equiv \half \intrrp \dfrac{\n(\br)\,\n(\br')}{|\br - \br'|}, \label{WH}
\een
while the exchange-correlation piece is:
\ben
W\xc[\n] \equiv 
\half \intr \n(\br) \intrp \dfrac{h\xc(\br,\br')}{|\br - \br'|}. \label{Wxc}
\een
Some approximations consider exchange and correlation separately,
often when hybridizing with Hartree--Fock \cite{Bb93,BEP97,HSE06}.  These approximations
use the exact exchange hole in whole or in part:
\ben
h\x(\br,\br') \equiv -\half \dfrac{| \gamma\s(\br,\br') |^2 }{ \n(\br) } \label{hx}
\een
where $\gamma\s(\br,\br')$ is the one-body reduced density matrix of the KS system:
\ben
\gamma\s(\br,\br') = 2\sum_{j=1}^{N/2} \phi_j^*(\br)\, \phi_j(\br') .
\een

While the foregoing allows one to evaluate the interaction XC energy
$W\xc[\n]$ for some given XC hole $h\xc(\br,\br')$, the kinetic correlation energy
$T\c[\n]$ cannot directly be found through the XC hole.
This is a consequence of using a non-interacting reference (i.e.\ the KS reference)
and its kinetic energy $T\s[\n]$.  
To get both $T\c[\n]$ and $W\xc[\n]$,
one must integrate from the KS system to the fully interacting system
using the adiabatic connection formalism \cite{LP75,GL76}.  This requires an infinite
number of fictitious systems, all with density $\n(\br)$, but whose electron-electron
repulsion is scaled by a factor of $\lambda$ called the coupling constant.  
The ground-state wavefunctions
$\Psi^\lambda[\n]$ thus minimize the expectation value of $\hatT + \lambda \hatW$
under the constraint of giving the density $\n(\br)$,
as in \Eqref{F}.
Then $\Psi^0[\n]$ is the KS wavefunction (likely a Slater determinant of the occupied KS
orbitals $\phi_j(\br)$), and $\Psi^1[\n]$ is the true wavefunction 
of the system \footnote{%
For systems with degeneracy, the coupling-constant wavefunctions $\Psi^\lambda$
should be replaced by coupling-constant mixed states $\Gamma^\lambda$.
}.  The full XC energy can then be found by an average of the XC
holes from coupling constant $\lambda =0$ to 1:
\ben
E\xc[\n] = \half \intr \n(\br) \intrp \dfrac{\bar h\xc(\br,\br')}{|\br - \br'|},
\een
where the coupling-constant averaged XC hole is:
\ben
\bar h\xc(\br,\br') = \int_0^1 d \lambda\, h^\lambda\xc(\br,\br').
\een
The XC hole at coupling constant $\lambda$, $h^\lambda\xc(\br,\br')$,
comes from the pair density $P^\lambda(\br,\br')$ of the
wavefunction $\Psi^\lambda[\n]$, as in \Eqref{pairhole}.  
Because $h\x^\lambda(\br,\br')$ comes from the KS orbitals (which are the same 
for all $\lambda$), $\bar h\x(\br,\br') = h\x(\br,\br') = h\x^\lambda(\br,\br')$.

\shd{Properties of the XC hole}
The XC hole has some simple properties due to 
the properties of $P^\lambda(\br,\br')$  \cite{B97b}.  The pair density is non-negative,
so 
\ben
h^\lambda\xc(\br,\br') \ge -\n(\br')\quad \forall\, \br,\br'; \label{holedepth}
\een
the pair density integrates over $\br'$ to yield $(N-1)\, \n(\br)$, which means that
\ben
\intrp h^\lambda\xc(\br,\br') = -1  \quad\quad\forall\, \br ; \label{holenorm}
\een
and utilizing the symmetry of the pair density (invariance under $\br \leftrightarrow \br'$), 
one can show:
\ben
\intrp \n(\br')\, h^\lambda\xc(\br',\br) = -\n(\br) \quad\quad \forall\, \br. \label{holedensity}
\een
These properties are ideal to build into models for $h\xc(\br,\br')$,
but many approximate functionals have XC holes which do not satisfy them.  
For example, 
the local density approximation (LDA) \cite{KS65} 
corresponds to the XC hole of the uniform electron gas \cite{GP02}.
The LDA XC hole is properly normalized 
\eqref{holenorm}, 
but does not always satisfy \Eqref{holedepth} (e.g.\ in certain circumstances when $\n(\br) > \n(\br')$). 
The spherical average $\bar h\xc(\br;u)$ of the
XC hole is defined as
\ben
\bar h\xc(\br;u) \equiv \intsphu \bar h\xc(\br,\br+u \hat{u}). \label{hxcsph}
\een
The LDA usually provides a good approximation of the short-range part (small $u$)
of the system average of \Eqref{hxcsph}, i.e. $n(\br)\, \bar h\xc(\br;u) $ 
integrated over all space \cite{BurPerLan-PRL-94,BurPer-IJQC-95,BurPerErn-JCP-98}.
If an XC hole is considered as a model of this spherical average, then
\Eqref{holedensity} cannot be directly assessed \cite{GorAngSav-CJC-09}; 
whether one thinks of LDA as an approximation to $h\xc(\br,\br')$ or $h\xc(\br;u)$ , therefore,
informs on whether LDA should satisfy \Eqref{holedensity} or not.
Regardless, the correct normalization of the LDA XC hole
(as in \Eqref{holenorm}) is a significant factor in the robustness of 
LDA \cite{JonGun-RMP-89}.
The forerunner to 
PBE \cite{Per-INC-91,PBE96} forced normalization of the XC hole
to cure problems with the gradient expansion of the density \cite{BP95},
giving insight into the need for the generalized gradient approximation (GGA).

\shd{Strictly correlated electrons}
The strictly correlated electron (SCE) functional 
\cite{Sei-PRA-99,SGS07,GorVigSei-JCTC-09, MalGor-PRL-12, MalMirCreReiGor-PRB-13}, 
corresponds to the $\lambda \to \infty$ 
limit of the adiabatic connection formalism,
completely opposite to the Kohn--Sham system at $\lambda = 0$.
The basic building blocks of the SCE method are called {\em co-motion} functions,
and are analogous to KS orbitals in KS-DFT.  Instead of minimizing the kinetic energy, however,
the co-motion functions minimize the interaction energy for a given density
\cite{S99,SGS07,MalGor-PRL-12, MalMirCreReiGor-PRB-13}.
The co-motion functions $\bff_j(\br)$ with $j=1,\ldots,N$ thus pinpoint classical locations
of the electrons.  Setting $\bff_1(\br) = \br$ to be the position of one electron,
we can write the interaction energy of the SCE method 
as \cite{SGS07,GorSeiVig-PRL-09,GorSei-PCCP-10}:
\ben
\label{Wsce}
W\sce[\n] \equiv \frac{1}{2}\intr \sum_{i= 2}^N \frac{\n(\br)}{|\br-\bff_i(\br)|},
\een
where the co-motion functions minimize this expression and satisfy
the two following physical constraints.

Because of the indistinguishability of  electrons, 
the co-motion functions must satisfy cyclic group properties.
Therefore knowledge of any non-trivial co-motion function $\bff_i(\br)$
is enough to generate all others:
\ben
\left.
\begin{array}{rcl}
\bff_1(\br) &\equiv& \br  \\
\bff_2(\br) &\equiv& \bff(\br) \\
\bff_3(\br) &=&  \bff(\bff(\br)) \\
&\vdots& \\
\bff_{N+1}(\br) &=& \underbrace{\bff(\bff(\ldots\bff(\bff(\br))))}_\text{$N$ times} = \br
\end{array}
\right\}.
\label{cyclic}
\een
Here we used $\bff_2(\br)$ as the
{\em co-motion generator} $\bff(\br)$ to produce the entire set.

In the SCE method, measuring the position of one electron also determines all others;
therefore the probability of finding an electron at position $\br$ must
be the same as finding an electron at any $\bff_i(\br)$ for $i=2,\ldots,N$.
Thus the co-motion generator $\bff(\br)$ must satisfy 
the nonlocal differential equation \cite{SGS07}:
\ben
\n(\br) = |J(\br)|\, \n( \bff(\br) ) ,
\label{eq_fi}
\een
where $J_{\mu \nu}(\br) = \partial f_{\mu}(\br) / \partial r_{\nu}$ are the Jacobian
matrix elements of $\bff(\br)$, and $|J(\br)|$ is the determinant.
Alternatively, \Eqref{eq_fi} can be expressed as an integral equation:
\ben
\int_{\Omega} d^3 r\, \n(\br) = \int_{\bff(\Omega)}\!\!\! d^3 r\, \n(\br), \label{equalintegral}
\een
where $\Omega$ is an arbitrary volume, and $\bff$ maps $\Omega$ 
to the volume $\bff(\Omega)$, i.e.\
\mbox{$\bff(\Omega) \equiv \{  \bff(\br) \ \forall\ \br \in \Omega  \}$}.

In 1d, we can find the co-motion functions as explicit functionals of the density, without performing
the minimization implicit in \Eqref{Wsce} \cite{ColDepMarNET,MalGor-PRL-12}.
For spherically symmetric 2d and 3d problems, the radial components of the co-motion functions can also be found
quite easily, while the angular components require minimizing the interaction energy over
the electronic angles \cite{SGS07,GorSei-PCCP-10,MenMalGor-PRB-14}.
But for a general 3d geometry, determining the co-motion functions is not as simple.  There is no shortcut;
the co-motion functions come out of $W\sce[\n]$ being minimized subject
to  constraints \eqref{cyclic} and \eqref{eq_fi}.   
There is an alternative approach to evaluate $W\sce[\n]$, 
the Kantorovich dual formulation \cite{ButDepGor-PRA-12,MenLin-PRB-13},
which bypasses the co-motion functions, and proves feasible for non-spherical systems.

It seems promising to develop approximations for the co-motion functions,
since they have a physically transparent meaning and role.
However, we proceed along somewhat different lines to develop our functional for strongly
correlated systems.

\sec{A nonlocal model of the XC hole for strong correlation}\label{newhole}

In this section we present our model for the XC hole in the strong-interaction limit, the 
nonlocal radius (NLR) XC hole, and describe the properties
of the resulting NLR energy functional and its functional derivative.
We present some non-self-consistent results with the NLR functional on exact atomic densities,
both real atoms and 1d pseudo-atoms, and compare to the SCE functional.  
The NLR model can be thought of as a real-space version of self-interaction correction \cite{PZ81,PRP14}, and 
it may be a descendant of the weighted density approximation \cite{AG77,AG78,GJL77,GJL79,DG90} on one
side and the PC model \cite{SPK00b,Ons-JPC-39} on the other; we discuss this likely ancestry after
introducing the concepts of the nonlocal NLR functional.
For further comparisons, we evaluate the NLR functional and others on the uniform electron gas.

The key ingredient for our NLR XC functional is a nonlocal generalization
of the Wigner--Seitz radius, inspired by 
work on orbital-free kinetic energy functionals \cite{H86}.  
We define this nonlocal Wigner--Seitz radius $R(\br)$ implicitly as the
radius of the sphere centered at $\br$ which encloses one electron:
\ben
\intrp \n(\br')\, \heavi\big(R(\br) - |\br' - \br| \big) \equiv 1, \label{Rdef}
\een
where $\heavi(x)$ is the Heaviside step function, equal to 0 for $x<0$ and 1 otherwise.
This is a simple generalization of the usual Wigner--Seitz radius, $r\s$,
which can be similarly defined in the uniform gas, using $\bu = \br' - \br$, $u = |\bu|$:
\ben
\intu \n\, \heavi\big( r\s - u) \equiv 1, 
\een
and which for non-uniform systems is typically generalized in a local way:
$r\s(\br) \equiv (3/(4 \pi \n(\br))^{1/3}$.  
The Wigner--Seitz radius has been used to
characterize the uniform electron gas from the beginning \cite{GiuVig-BOOK-05},
since it quantifies an effective distance between electrons.
The nonlocal generalization $R(\br)$  reduces to the local radius $r\s(\br)$ for
uniform systems, but it offers greater physical insight about the average
number of electrons near each point in a non-uniform system.

We now use the nonlocal information contained
within the nonlocal Wigner--Seitz radius $R(\br)$ to design
a model XC hole which is
correct for one-electron-like
systems, e.g.\ in dissociating H$_2$, as well as one-electron systems.
Our XC hole sets the pair density to zero for electron coordinates within
the nonlocal radius:
\ben
h\xc\NL(\br,\br') = -\n(\br')\, \heavi\big( R(\br) - |\br' - \br| \big).\label{nlhole}
\een
This NLR XC hole models
systems in which the electron wavefunction allows no
two electrons to get close to each other, i.e., systems which are strongly correlated.
This means that $h\xc\NL(\br,\br')$ is an approximation for the $\lambda \to \infty$ limit
of the XC hole -- the SCE hole -- and not the coupling constant averaged hole $\bar h\xc(\br,\br')$.
In much the same way, 
SCE physics describes the situation in which each electron excludes the others from a volume
in which the density integrates to 1.  
Therefore, if we approximate $\bar h\xc(\br,\br')$ by $h\xc\NL(\br,\br')$, we
expect to obtain energies far too low for most chemical systems, just like in KS-SCE
\cite{MalGor-PRL-12,MalMirGieWagGor-PCCP-14}.

For efficiency of notation, 
it is convenient to define the volume $\Omega(\br)$ over which the Heaviside step
function of \eqref{Rdef} is non-zero.  The volume $\Omega(\br)$
is defined as the sphere centered at $\br$ with radius $R(\br)$, so that we can rewrite
\Eqref{Rdef} as:
\ben
\intOrp \n(\br') \equiv 1.
\een
Now plugging \Eqref{nlhole} into \Eqref{Wxc},
we find the NLR interaction XC energy to be:
\ben
W\xc\NL[\n] \equiv -\half \intrOrp \dfrac{\n(\br) \, \n(\br')}{|\br - \br'|},\label{WxcNL}
\een
where the $\br'$ integral only integrates over the nonlocal volume $\Omega(\br)$.
Thus the hole of \Eqref{nlhole}
completely removes the Hartree interaction between the density at $\n(\br)$ and $\n(\br')$
if $\br'$ is within the sphere $\Omega(\br)$ centered at $\br$.

\begin{figure}
\includegraphics[width=\columnwidth]{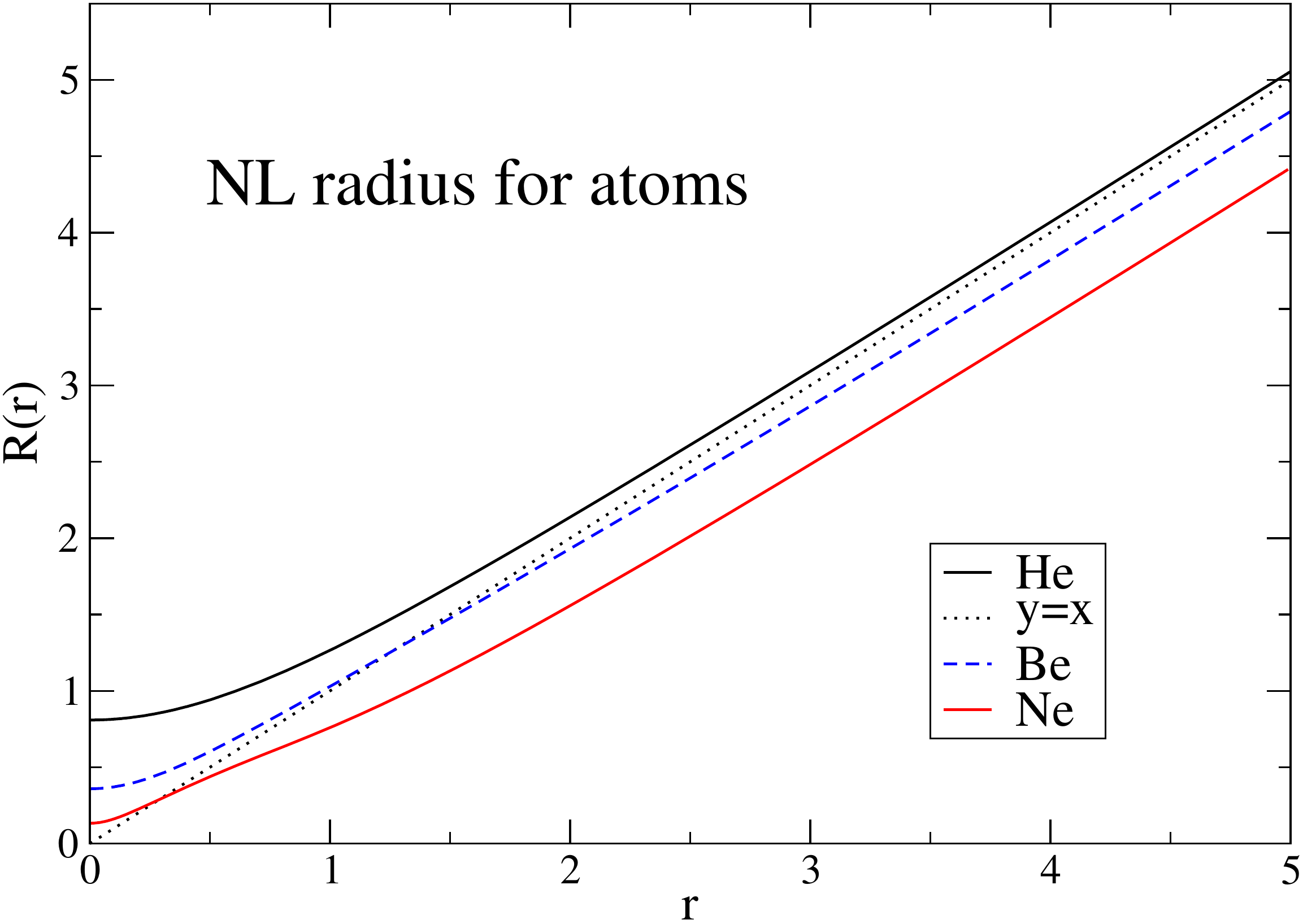}
\caption{%
Nonlocal radius $R(r)$ \eqref{Rdef} for the exact densities 
of the spherically symmetric atoms helium, beryllium, and neon; exact data from \Refs{UG94,FGU96,ARU98}.
To see the asymptotic behavior, the $y=x$ line is also plotted.
}
\label{3datomR}
\end{figure}

\begin{figure}
\includegraphics[width=0.85\columnwidth]{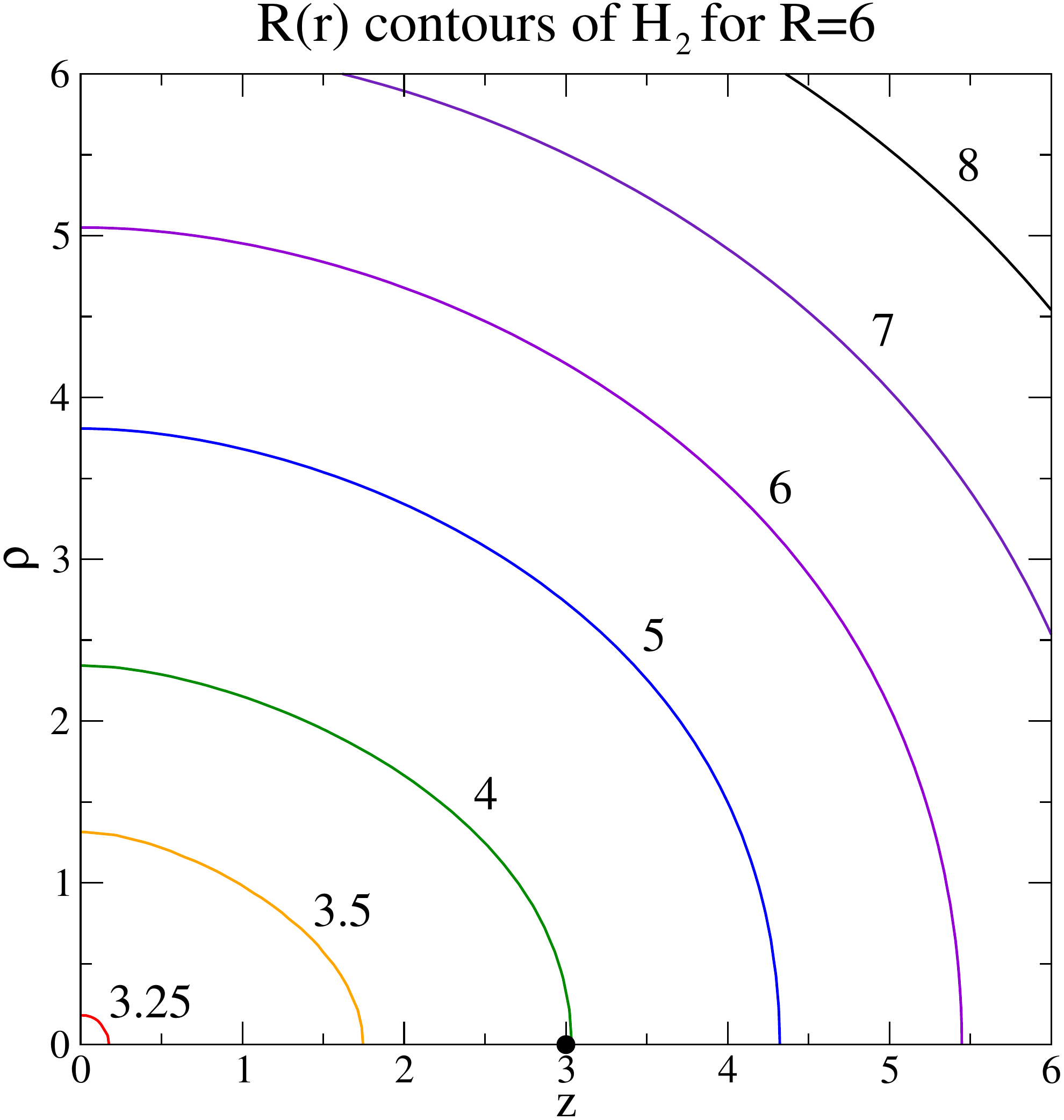}
\caption{%
Nonlocal radius $R(\br)$ \eqref{Rdef} for the exact density of the H$_2$ molecule with bond length $R=6$,
in cylindrical coordinates $(z,\rho)$.
Exact density is the Full-CI result from GAMESS-US \cite{GAMESS} within the aug-cc-pV6Z \cite{aug-cc-pVXZ}
basis set.  The bond-axis is along the
$z$ coordinate, and the nuclei are at $z = \pm 3$.
}
\label{3dH2R}
\end{figure}

To give an idea of what is going on inside the NLR XC model,
we discuss some properties of the nonlocal (NL) radius $R(\br)$.
For $N\le 1$ systems, $R(\br) \to \infty$, so that the NLR interaction XC energy
cancels the Hartree energy, i.e.\ $W\xc\NL[\n] = -W\H[\n]$.  For all other (finite) systems,
the NL radius asymptotically goes like $R(\br) \to r - c(\hat r)$ as $r \to \infty$, 
where $c(\hat r)$ depends on how many
electrons there are, as well as the direction of $\br$ for
non-spherical systems.  
We can easily calculate $R(\br)$ for any given density by fitting
the density to a sum of exponentials or Gaussians, and this is explained in \Appref{calculateR}.
In \Figref{3datomR}, we show $R(\br)$ for a few atoms.
The NL radius has bumps and curves due to shell structure,
but these are rather gentle since $R(\br)$ is defined by an integral over the density. 
Asymptotically, $c(\hat r) = 0$ for helium, and more generally $c(\hat r) = 0$ for
any spherically symmetric $N=2$ system.  The next leading term in $R(\br)$,
a $1 / r$ term with some small coefficient, 
explains why $R(\br)$ for helium does not look yet like $r$ 
for the larger $r$ values in \Figref{3datomR}.
In \Figref{3dH2R} we show a contour
plot of $R(\br)$ for the hydrogen molecule at bond length $R=6$.  Due to a lack of shell
structure in H$_2$, $R(\br)$ is rather bland and featureless.  Its contours in the 
$(z,\rho)$ plane are roughly ellipses which tend towards circles at large distances.

\shd{Functional derivatives}
We can put $W\xc\NL[\n]$ into a symmetric form:
\ben
W\xc\NL[\n] = \half \intrrp \dfrac{\n(\br)\, \n(\br')}{|\br-\br'|} \, g\xc\NL(\br,\br')\label{WxcNLsym}
\een
where the NLR pair XC function is:
\ben
g\xc\NL(\br_1,\br_2) \equiv - \half \Big( \heavi\big( R(\br_1)- r_{12} \big) + \heavi\big( R(\br_2)-r_{12}  \big) \Big),\label{gxcNL}
\een
with $r_{12} \equiv |\br_1 - \br_2|$.  
The functional derivative can then be written with this pair XC function as:
\bea
v\xc\NL[\n](\br)  &=& \intrp \dfrac{\n(\br')}{|\br-\br'|}\,  g\xc\NL(\br,\br') \nonumber \\
    && +\half \intrp \dfrac{ n(\br') }{R(\br')}\, \heavi\big( R(\br') - |\br-\br'|), \ \  \ \label{vxcNL}
\eea
and see \Appref{calculatev} for a full derivation.
The second term integrates over all points $\br'$ which are within their own
Wigner--Seitz radius $R(\br')$ of the point $\br$.  We call this region the
NL reflection volume $\Iomega(\br)$,
and its definition is similar to the NL volume $\Omega(\br)$:
\ben
\left.
\begin{array}{rcl}
\Omega(\br) &\equiv& \{ \br'\, :\,  R(\br) > |\br - \br'| \} \\[5pt]
\Iomega(\br) &\equiv& \{ \br'\, :\,  R(\br') > |\br - \br'| \} 
\end{array}
\right\}.
\een
There is another such $\Iomega(\br)$ integration region in the first term of \Eqref{vxcNL}
due to the NLR pair XC function.
These integrals are straightforward, albeit numerically challenging, to evaluate.

By construction, the NLR XC hole satisfies the correct XC hole normalization
in \Eqref{holenorm}, in a nonlocal and physically meaningful way.
In addition, the nonlocal XC hole satisfies another constraint on the exchange-correlation hole:
\Eqref{holedepth}, since $h\xc\NL(\br,\br') \ge -\n(\br')$.
But since $h\xc\NL(\br,\br')$ does not satisfy \Eqref{holedensity}, 
$v\xc\NL[\n](\br)$ does not go like $-1/r$ as $r \to \infty$ as it should.
Instead, $v\xc\NL[\n](\br) \to -1/(2r)$ as $r\to\infty$.  One can see this
by examining \Eqref{gxcNL} and \Eqref{vxcNL}.
As $r \to \infty$, one step function
inside $g\xc\NL(\br,\br')$  will vanish---the term with $R(\br')$---since $R(\br')$ is rather small
near the molecular center.  Integrating the density at $\n(\br')$ with the other step function yields
exactly one electron, but with  minus one-half out front (in \Eqref{gxcNL}) and the
Coulomb operator inside the integral in \Eqref{vxcNL}, the result is $v\xc\NL[\n](\br) \to -1/(2 r)$.

\shd{Non-self-consistent results for atoms and molecules}
In \Tabref{NLatoms},
we evaluate $W\xc\NL[\n]$ for the exact densities of simple atoms and compare to
SCE results and the exact $E\xc[\n]$.  As expected, for chemical systems
$W\NL\xc[\n]$ is too low to approximate $E\xc[\n]$.  A generalized-gradient approximation
to the PC model, PC-GGA \cite{SPK00b}, is also tabulated for these atoms.  We discuss the PC model
later in terms of NLR quantities, but here we note that its GGA incarnation $W\xc\PCGGA[\n]$
behaves energetically quite similarly to $W\xc\NL[\n]$, as does $W\xc\SCE[\n]$.
These strongly correlated methods are all lower than the exact $E\xc[\n]$, but
the nonlocal methods (KS-NLR and KS-SCE) are exact for one-electron systems.
However, for H, PC-GGA does very well:  the exact $E\xc[\n] = -0.312500 = W\xc\NL[\n] = W\xc\SCE[\n]$, 
while $W\xc\PCGGA[\n] = -0.312767$, an error of less than 0.1\%. 
Notice however, that while energies can be very similar when the functionals 
are evaluated on accurate densities, the functional derivatives (potentials) 
behave very differently. For example, charge localization without magnetic 
order is obtained self-consistently by KS-SCE
\cite{MalMirCreReiGor-PRB-13,MenMalGor-PRB-14} and by KS-NLR (see \Secref{results}), 
while it is missed by any local or semilocal functional. 

\begin{table}
\begin{tabular*}{\columnwidth}{@{\extracolsep{\fill}}lcccc}
\hline
atom            & $E\xc[\n]$ 
                        & $W\NL\xc[\n]$   
                                 & $W\xc\PCGGA[\n]$ 
                                         & $W\xc\SCE[\n]$  \\
\hline
H$^-$           & $-$0.423 & $-$0.543 & $-$0.555 & $-$0.569   \\
He              & $-$1.067 & $-$1.426 & $-$1.463 & $-$1.500    \\
Li              & $-$1.799 & $-$2.496 & $-$2.556 & $-$2.603     \\
Be              & $-$2.770 & $-$3.835 & $-$3.961 & $-$4.021    \\
Ne              & $-$14.49 & $-$18.28 & $-$20.00 & $-$20.04     \\
\hline
\end{tabular*}
\caption{
Evaluating the NLR functional on various exact 3d atomic densities,
and comparing against 
exact numbers \cite{CLB98,FGU96,UG94,SSP86}, the generalized gradient approximation (GGA) of the PC
model \cite{SGS07}, as well as the SCE model \cite{SGS07,MSG12}.
Exact densities from \Refs{UG94,SGS07,FGU96,ARU98} 
are fitted to a sum of exponentials to evaluate
$R(\br)$ and thus $W\xc\NL[\n]$ as described in \Appref{calculateR}.
Lithium $E\xc[\n]$ is done in a pure DFT way---i.e.\ with a set of spin-restricted KS orbitals---using
the KS potential of \Ref{SGS07} and the CCSD(T)=FULL results of the CCCBDB \cite{CCCBDB}
using the aug-cc-pVQZ basis set \cite{aug-cc-pVXZ}.
}
\label{NLatoms}
\end{table}

\begin{figure}
\includegraphics[width=\columnwidth]{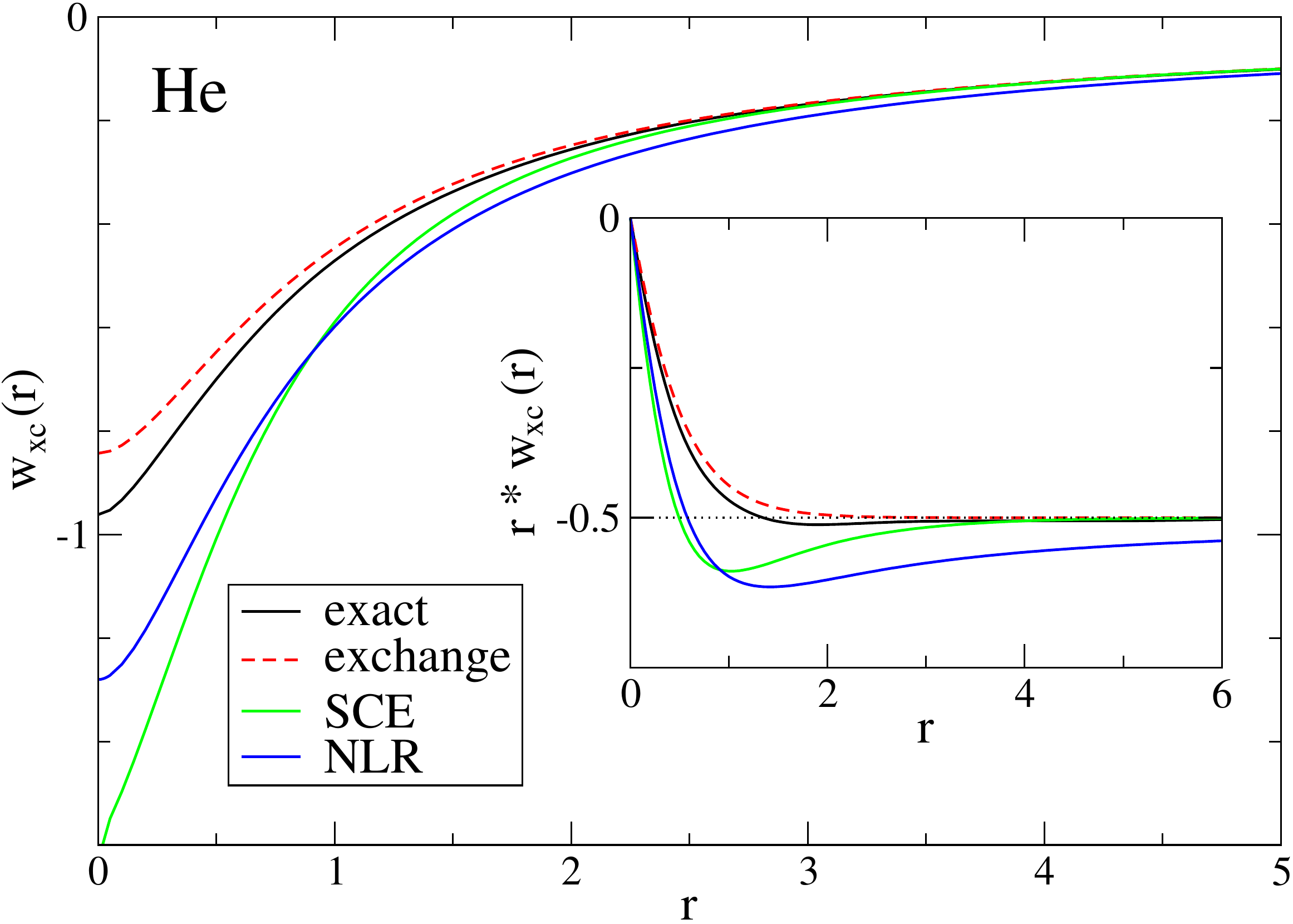}
\caption{%
The interaction XC  energy per particle \eqref{wxcholegauge} as a function of the radius $r$
for the exact 3d helium atom density \cite{MSG12}.  
Inset:  $r\,w\xc(\br)$ to see
asymptotic behavior.  The exact $w\xc(\br) \to -1/(2r)$ as $r\to\infty$.
}
\label{3dHewxc}
\end{figure}

\begin{figure}
\includegraphics[width=\columnwidth]{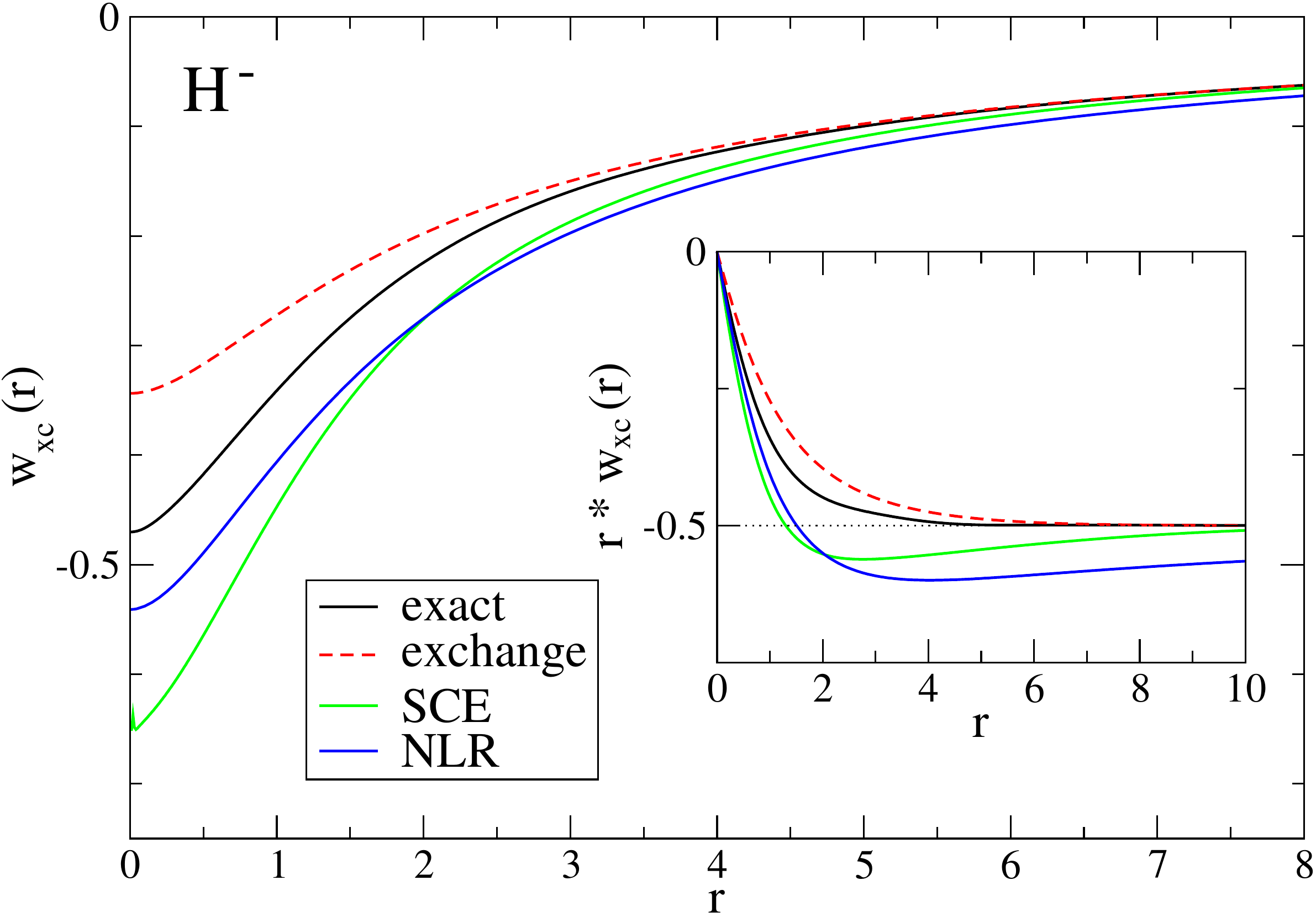}
\caption{%
The interaction XC energy per particle \eqref{wxcholegauge} as a function of the radius $r$
for the exact 3d H$^-$ atom density \cite{MSG12}.  
Inset:  $r\,w\xc(\br)$.
}
\label{3dHmwxc}
\end{figure}

\begin{figure}
\includegraphics[width=\columnwidth]{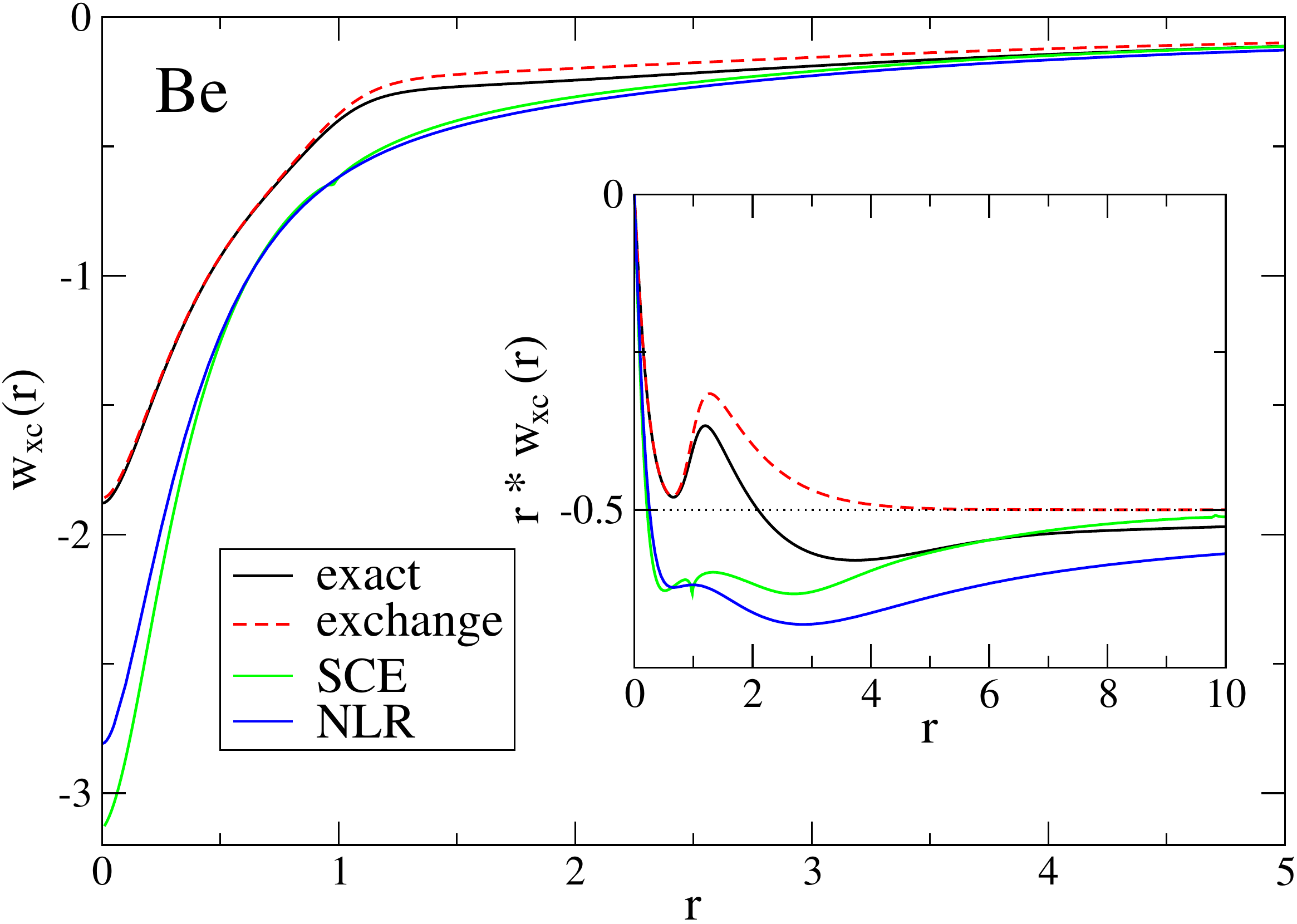}
\caption{%
The interaction XC  energy per particle \eqref{wxcholegauge} as a function of the radius $r$
for the exact 3d beryllium atom density \cite{MSG12}. 
Inset:  $r\,w\xc(\br)$.
Notice that the SCE energy density physically has a kink near $r=1$, which has to do with
classical electrons disappearing to infinity in the SCE method \cite{MSG12}.
}
\label{3dBewxc}
\end{figure}

As in other work \cite{MSG12}, we compare interaction XC  energy densities
to understand the properties of the nonlocal functional.
We define the interaction XC energy per particle $w\xc[\n](\br)$ as:
\ben
w\xc[\n](\br) \equiv \half \intrp \dfrac{h\xc(\br,\br')}{|\br - \br'|}, \label{wxcholegauge}
\een
so that $W\xc[\n] = \intr \n(\br)\, w\xc[\n](\br)$.  Like all energy densities, 
$w\xc[\n](\br)$ has a gauge, since the introduction of the Laplacian of any function,
i.e.\ $w\xc[\n](\br) \to w\xc[\n](\br) + (\nabla^2 f[\n](\br)) / \n(\br)$, will give the same
integral $W\xc[\n]$ \cite{BCL98}.  However, in \Eqref{wxcholegauge}, 
we haven chosen the XC
hole gauge where $w\xc[\n](\br) \to w\x[\n](\br) \to -1/(2 r)$ as $r \to \infty$.
In this gauge, the exchange energy per particle is (using \Eqref{hx}):
\ben
w\x[\n](\br) = -\sum_{i,j=1}^{N}\dfrac{\phi_i^*(\br)\, \phi_j(\br)}{\n(\br)} \intrp \dfrac{\phi_j^*(\br')\,\phi_i(\br') }{|\br - \br'|}, \label{wx}
\een
the SCE interaction energy per particle becomes \cite{MSG12}:
\ben
w\xc\SCE[\n](\br) \equiv \half \sum_{i=2}^N \dfrac{1}{|\br - \bff_i(\br)|} - \half v\H[\n](\br), \label{wxcSCE}
\een
and the NLR interaction energy per particle is:
\ben
w\xc\NL[\n](\br) = -\half \intOrp \dfrac{\n(\br')}{|\br - \br'|}. \label{wxcNL}
\een
We plot these energy densities for various atoms:  helium in \Figref{3dHewxc}, 
the hydrogen anion in \Figref{3dHmwxc}, and beryllium in \Figref{3dBewxc}.
For these systems, the exact $w\xc(\br)$ lies below the exchange-only 
$w\x(\br)$, above
$w\xc\NL(\br)$, and usually above $w\xc\SCE(\br)$.
For beryllium at large $r$, however, the reverse occurs \footnote{That $w\xc\SCE(\br)$
can sometimes be above the exact $w\xc(\br)$ is, while unusual, not too surprising.  This
has also been observed for the Hookium atom \cite{MSG12}.  Basis set dependence is strong
out in the tail region of these atoms, but to the scale of the figures we believe
we are converged to the basis set limit.}.
One flaw of the NLR method is that its interaction XC energy per particle
decays quite slowly to the correct $-1/(2r)$ behavior.

With the above properties, $w\xc\NL(\br)$ may be well suited
as an ingredient to approximate the true $w\xc(\br)$
using a local-weighting approach \cite{MSG12,AK08}:
\ben
w\xc\HYB(\br) \equiv \dfrac{\alpha(\br)\, w\xc\NL(\br) + w\x(\br)}{\alpha(\br) + 1},
\een
where $\alpha(\br)$ becomes large in regions where strong correlation effects
are important, and goes to zero where HF is sufficient.
While we will not pursue
this idea any further in this work, we remark here that this may be seen as
integrating $h\xc^\lambda(\br,\br')$ to remove $\lambda$ dependence in the adiabatic
connection formalism using some local (or nonlocal) information 
at position $\br$ \cite{MSG12,AK08}.  
This local weighting is inspired by the interaction strength interpolation (ISI) method, which
obtains $E\xc[\n]$ directly by integrating a model of $W\xc^\lambda[\n]$ 
\cite{Ern-CPL-96,S99,SPK00,SPK00b}.
The advantage of local weighting is that it is inherently size-consistent, whereas
the ISI is not \cite{MSG12}.

\begin{figure}
\includegraphics[width=\columnwidth]{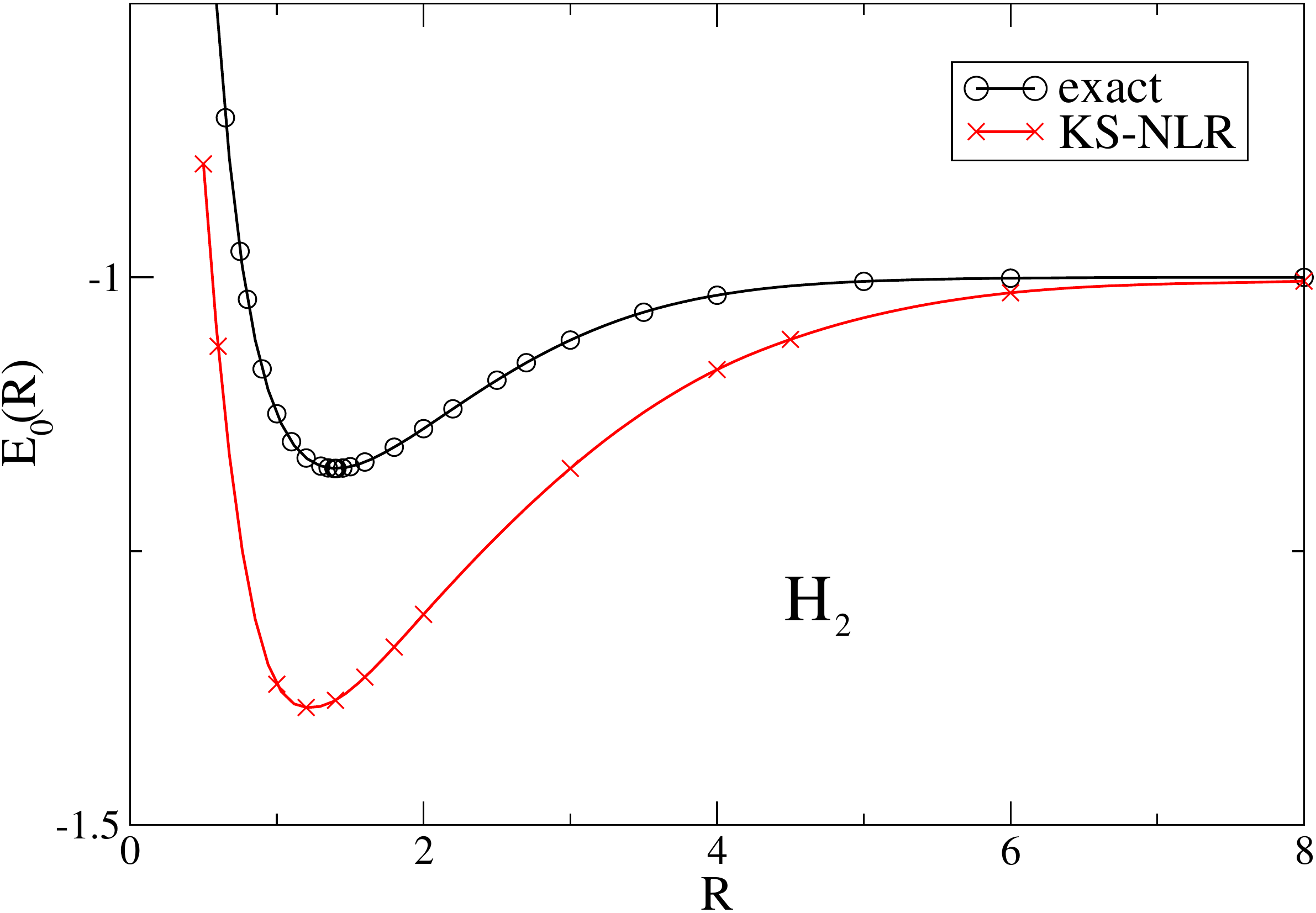}
\caption{%
3d H$_2$ comparing exact energies from \Ref{KW68} with non-self-consistent KS-NLR energies
on exact densities; densities from
GAMESS-US \cite{GAMESS} Full-CI calculations within the aug-cc-pV6Z \cite{aug-cc-pVXZ} 
basis set, with fits thanks to 
Stefan Vuckovic.  
Lines are cubic spline interpolations
through data (at markers).
}
\label{3dH2}
\end{figure}

We are also able to calculate H$_2$ within the KS-NLR method, which we present here non-self-consistently.
This calculation is much more difficult for KS-SCE, due to the lack of a general 3d geometry solver.
With KS-NLR, H$_2$ is rather straightforward, though the challenge is finding $R(\br)$.  
As already seen in \Figref{3dH2R}, $R(\br)$ is a rather simple function, so we use a very
simple grid exploiting the cylindrical symmetry of H$_2$.  The result is in \Figref{3dH2}.
We expected KS-NLR to be a lower bound to the energy, and it is clear from the figure that
KS-NLR does poorly except for large bond distances (at very strong correlation).  Nevertheless,
the figure confirms what we would expect from KS-SCE considering 1d results 
\cite{MalMirGieWagGor-PCCP-14} (and see later
in this paper).

\shd{Ancestry and relatives}
In his original approximation, 
Hartree corrects for self-interaction within the orbitals \cite{H28}:
\ben
W\xc\Hartree \equiv
 -\half \sum_\sigma \sum_{i=1}^{N_\sigma} \intrrp \dfrac{|\phi_{i\sigma}(\br)|^2 \, |\phi_{i\sigma}(\br')|^2}{|\br - \br'|},\label{HartreeWxc}
\een
where we include spin indices $\sigma$ for the orbitals.  For $N=2$ electrons, this approximation
is equivalent to Hartree--Fock, and with spin-symmetry-breaking it is unrestricted Hartree--Fock.
For $N>2$ electrons, this approximation does not include exchange effects, however, and it is not 
invariant under unitary orbital rotations.
In contrast, the NLR interaction XC  energy $W\xc\NL[\n]$ {\em is} invariant under unitary transformations of the orbitals, since it only depends on the density (and the NL radius which can be found from the 
density).
The NLR method can thus be thought of as a real-space version of self-interaction correction.
One may compare more recent orbital-based corrections
that are either non-unitary \cite{PZ81} or unitary \cite{PRP14},
but look less like the NLR interaction XC energy
than \Eqref{HartreeWxc}.

Due to including some nonlocal information about the density,
the NLR functional is also similar in spirit to the weighted density approximation (WDA) \cite{AG77,AG78,GJL77,GJL79,DG90},
which uses the pair XC function of the uniform electron gas, but with
an averaged version of the density to enforce the correct hole normalization.
The functional is designed to be correct in the uniform gas, but have an improved hole for non-uniform systems.
Recent WDA's yield results of similar quality to GGA functionals (at a higher computational cost), 
but with the advantage of avoiding symmetry breaking (see, e.g., \Ref{CCRC12}).

Finally, the NLR functional appears to be most closely related to
the point charge plus continuum (PC) model \cite{SPK00b,Ons-JPC-39}.
The PC model finds inspiration from Wigner's treatment \cite{W34} of the strongly correlated, very-low-density 
($r\s \gtrsim 100$)
uniform electron gas, for which he was able to obtain accurate correlation energies.
In the PC model \cite{Ons-JPC-39}, the XC energy is obtained as the electrostatic energy
of an electronic system in a fictitious positive background  with the same density $n(\br)$.
Specifically, one determines the electrostatic energy of a given cell $\Omega(\br)$,
wherein the density integrates to one.  
(Previous work on the PC model uses local or semilocal approximations to $\Omega(\br)$
\cite{SPK00b}, but for a closer comparison we use the nonlocal $\Omega(\br)$.)
The cell is modeled as
one point-charge electron (at $\br$) and the positive background in the volume $\Omega(\br)$.
The cell energy is then the sum of the interaction of the point charge at $\br$ 
with the positive background and the background-background interaction in the 
cell.  One then sums over and averages the energy of the cells \cite{SPK00b,MSG12}:
\bea
 {W}\xc\PC[\n] &=& -\intr\n(\br) \intOrp \dfrac{\n(\br')}{|\br - \br'|} \label{WxcPC} \\
&&+ \half \intr \n(\br) \intOrp \!\intOrrp \dfrac{\n(\br')\, \n(\br'')}{|\br' - \br''|}. \nonumber
\eea
Comparing \Eqref{WxcPC} and \Eqref{WxcNL},
one might roughly think of $W\xc\NL[\n]$ as some approximation to $W\xc\PC[\n]$---if 
each integral (without its coefficient) has about the same magnitude.

In \Tabref{PCNL1d},
we verify for some simple 1d pseudo-atoms and molecules that using
the nonlocal radius to define $\Omega(\br)$ for the PC model results in energies very
close to the nonlocal model, though neither is consistently closer to the exact $E\xc[\n]$.   
In 1d we use soft-Coulomb interactions between
electrons and nuclei:  $w(u) = 1/\sqrt{ u^2 + 1 }$, with appropriate coefficients
due to nuclear charges.
\Ref{WSBW12} explains the methodology for both
exact and approximate solutions to these 1d systems;
such model systems are analogous to simple 3d systems and allow quick
evaluation and prototyping of functionals.  The 1d H$_2$ data in \Tabref{PCNL1d} suggest
that the PC model as well as the SCE model energies for the 3d H$_2$ molecule
would be a little deeper and asymptotically slower to converge to the isolated
atom limit than KS-NLR in \Figref{3dH2}.  This may not be true for other molecules, however,
which we will see later when considering 1d LiH.  Despite the small advantage to KS-NLR for H$_2$, 
we reiterate that all of these nonlocal functionals give a very low estimate for $E\xc[\n]$.

\begin{table}
\begin{tabular*}{\columnwidth}{@{\extracolsep{\fill}}lcccc}
\hline
system & $E\xc[\n]$ 
               & $W\NL\xc[\n]$   
                        & $W\xc\PC[\n]$ 
                                & $W\xc\SCE[\n]$  \\
\hline
1d H$^-$& -0.595 & -0.747 & -0.727 & -0.756 \\
1d He   & -0.733 & -0.877 & -0.871 & -0.889 \\
1d Li   & -1.087 & -1.288 & -1.275 & -1.303 \\
1d Be   & -1.481 & -1.782 & -1.788 & -1.818 \\
1d eq.~H$_2$& -0.683 & -0.836 & -0.838 & -0.846 \\
1d str.~H$_2$ & -0.661 & -0.700 & -0.717 & -0.713 \\
\hline

\end{tabular*}
\caption{
Evaluation of NLR \eqref{WxcNL}, PC (with the NLR $\Omega(\br)$) \eqref{WxcPC}, 
and SCE interaction XC functionals on the exact
density of 1d systems given
soft-Coulomb interactions between electrons.  
Exact data and densities from \Ref{WSBW12}.
Equilibrium (eq.) and stretched (str.) 1d H$_2$ are with bond lengths $R=1.6$ and $5$, respectively \cite{WSBW12}.
}
\label{PCNL1d}
\end{table}

We mention here that
due to the triple integral in \Eqref{WxcPC}, the PC model would be {\em much} 
more expensive to implement self-consistently than the NLR functional.
Unlike in the NLR model, the functional derivative of $W\xc\PC[\n]$ does not simplify nicely,
and one is left with single-, double-, and triple-integrals 
just to evaluate the potential at one point, $v\xc\PC[\n](\br)$.  This is true
even if a local approximation to $R(\br)$ is used, unless a local approximation is also applied to the integrals.
Doing this, however, would destroy some of the nice properties of the PC model.

As a final remark on the PC model, we
notice that if one uses the nonlocal $R(\br)$ (and $\Omega(\br)$) in the PC model,
the corresponding XC potential has the right asymptotic behavior.
In fact, the second term (the background self-interaction term) in \Eqref{WxcPC} is short-ranged,
and the first term is twice $W\xc\NL[\n]$.  Since the long-ranged part of $v\xc\NL[\n](\br)$ goes like
$-1/(2r)$, the long-ranged part of $v\xc\PC[\n](\br)$ is thus twice that, or $-1/r$.
This behavior is interesting and warrants further research.

\shd{The uniform electron gas}
There is one many-body system where we can analytically investigate the
behavior of these strong correlation functionals:  the uniform electron gas,
studied from days of yore \cite{W34,GiuVig-BOOK-05}. 

In the uniform gas, $R(\br) \to r\s$, so we can easily calculate the per-particle energy
of $W\xc\NL[\n]$ as a function of $r\s$:
\bea
w\xc\NLunif(r\s) &=& -\half \intu \n\, \theta( r\s - u ) / u \nonumber \\
&=& -2\pi\n \int_0^{r\s} du\, u \nonumber \\
&=& -\pi\n r\s^2  \nonumber \\
&=& -0.75 / r\s \label{wxcNLunif}
\eea
The local approximation to the NLR XC energy would thus be:
\ben
W\xc\NLunif[\n] \equiv \intr \n(\br)\, w\xc\NLunif\big(r\s(\br)\big). \label{WxcNLDA}
\een
The SCE functional gives
the correct interaction XC energy 
for the strongly correlated (or low density) uniform
electron gas \cite{MSG12}.  
The interaction XC 
energy per particle in this $r\s\to\infty$ limit is \cite{SPK00b,GiuVig-BOOK-05}:
\ben
w\xc\SCEunif(r\s) \approx -0.89593 / r\s. \label{wxcSCEunif}
\een
The PC interaction XC  energy per particle has also been calculated
for the uniform gas \cite{SPK00b} 
\ben
w\xc\PCunif(r\s) = -0.9 / r\s 
\een
which is quite close to the exact low-density limit (see \Ref{MSG12} for further discussion).
We emphasize here that both $W\xc\NL[\n]$ and $W\xc\PC[\n]$ are approximations to the
strongly correlated limit, so they both make an error on the low-density uniform gas.
However, the Wigner crystal is achieved only in the ultra-low density regime,
around $r\s \gtrsim 100$.  For perspective, $r\s$ is about 100 at a distance of
7 bohr radii from a hydrogen atom,
so these limits may not be too useful in practice.

There is one observation that we should make given the above.
Even though KS-NLR will give low energies for many systems of chemical interest,
it will not necessarily yield a lower bound to the energy of any system (contrary to KS-SCE, which is guaranteed to yield a rigorous lower bound to the exact energy);
in the ultra-low density uniform electron gas we evidently have $E_v\KSNL[\n] > E_v[\n]$.

As we proceed, we leave behind the local $W\xc\NLunif[\n]$ \eqref{WxcNLDA}.
Using a local approximation in our interaction XC energy obviously nullifies
the interesting nonlocal physics of the NLR functional,
including its ability to capture one-electron and one-electron-like 
systems correctly.
However, the uniform gas may have a different role to play when building corrections to KS-NLR, which we will address
in future work.

\sec{Self-consistent nonlocal results}\label{results}

In this section, we use $W\xc\NL[\n]$ to approximate $E\xc[\n]$ in the KS framework,
and run self-consistent calculations for a few systems where it is also possible to compare to KS-SCE
and exact results.
We find that in many systems 
the KS-NLR functional behaves like the KS-SCE functional, differing only somewhat for anions.

The energy in the KS-NLR method is:
\ben
E_v\KSNL[\n] \equiv T\s[\n] + \intr \n(\br)\, v(\br) + W\H[\n] + W\xc\NL[\n]. \label{EvKSNL}
\een
To perform self-consistent calculations with this functional,
we must include $v\xc\NL[\n](\br)$ in the KS potential:
\ben
v\s\NL(\br) = v(\br) + v\H[\n](\br) + v\xc\NL[\n](\br).
\een
Asymptotically,
since $v\xc\NL[\n](\br) \to -1/(2r)$ as $r\to\infty$, we have:
\ben
v\s\NL(\br) \to \dfrac{N-Z - \frac{1}{2}}{r}\quad\quad (r\to\infty),
\een
where $Z$ is the total charge of all nuclei in the system.  The correct
asymptote of $v\s(\br)$, however, is $(N-Z-1)/r$, which is very important
in anions.  
Many standard LDA and GGA functionals have short-ranged $v\xc(\br)$, however,
so their asymptotic behavior is worse than the NLR functional.  The exception is
the B88 functional \cite{B88}, which like KS-NLR functional has half the right asymptotic behavior:
$v\xc(\br) \to -1/(2 r)$.
For B88, this is enforced using the exponential decay of the density, however, which is true only for
atoms and molecules.
Within parabolic traps, therefore, the B88 functional needs modifications \cite{VRMP14}, 
while the NLR functional does not.  
The KS-SCE method is also density-decay indifferent,
but its XC potential goes correctly to $-1/r$ as $r\to\infty$.

For our 3d results, we perform self-consistent atomic calculations 
by diagonalization on a simple radial grid (spherically averaging the density).
We numerically integrate for the nonlocal radius $R(r)$ as well as the
NLR XC potential $v\xc\NL(r)$ each iteration.  This simple \mbox{\textsc{NLRatoms}}
code is freely available online \footnote{%
The basic KS-NLR code for atoms can be found at
\href{https://github.com/lowagner/NLRatoms}
{https://github.com/lowagner/NLRatoms}.
}.  
In 1d, we use the machinery of \Refs{WSBW12,MalMirCreReiGor-PRB-13} to self-consistently determine
the energies of pseudo-atoms and molecules.

\ssec{3d atoms}

In this section we study real 3d atoms.
We start with $N=2$ (helium and hydrogen anion atoms),
since KS-NLR gives the correct energies for $N=1$ systems.
We study the challenge of binding the hydrogen anion in KS-DFT
by two methods:  variable $N$ from 0 to 2 with
$Z=1$, and fractional $Z$ with fixed $N=2$.
Finally we consider ionization energies and the HOMO eigenvalues
of the KS-NLR method for small atoms up to neon.

\begin{figure}
\includegraphics[width=\columnwidth]{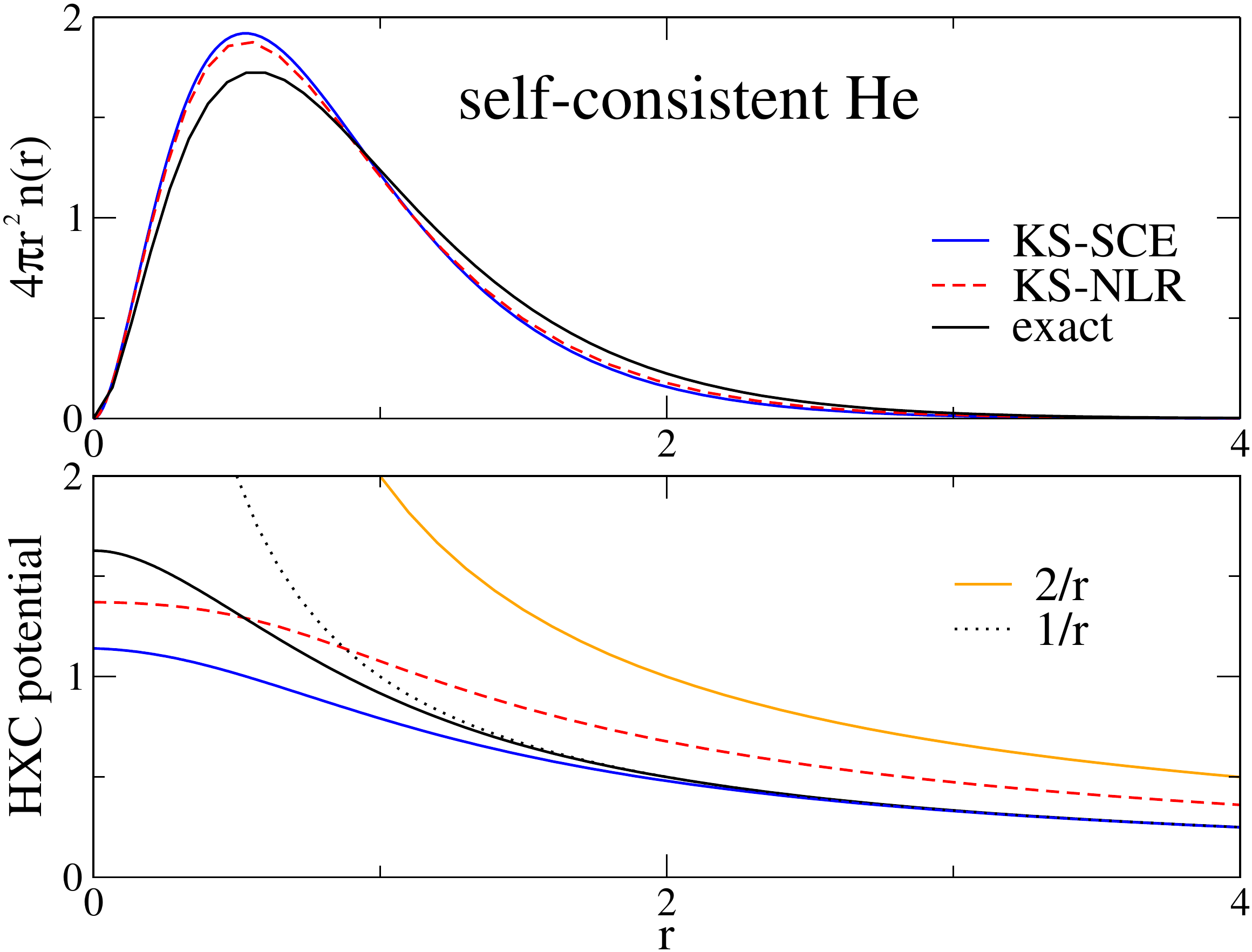}
\caption{%
Self-consistent results for the KS-SCE and KS-NLR \eqref{EvKSNL}  energy functionals
of the 3d helium atom:  plotted are the densities and HXC potentials, with comparisons
to $1/r$ ($2/r$), the asymptotic behavior of the exact (LDA) HXC potential.
Exact results thanks to Cyrus Umrigar \cite{UG94}.
}
\label{3dHe}
\end{figure}

\begin{figure}
\includegraphics[width=\columnwidth]{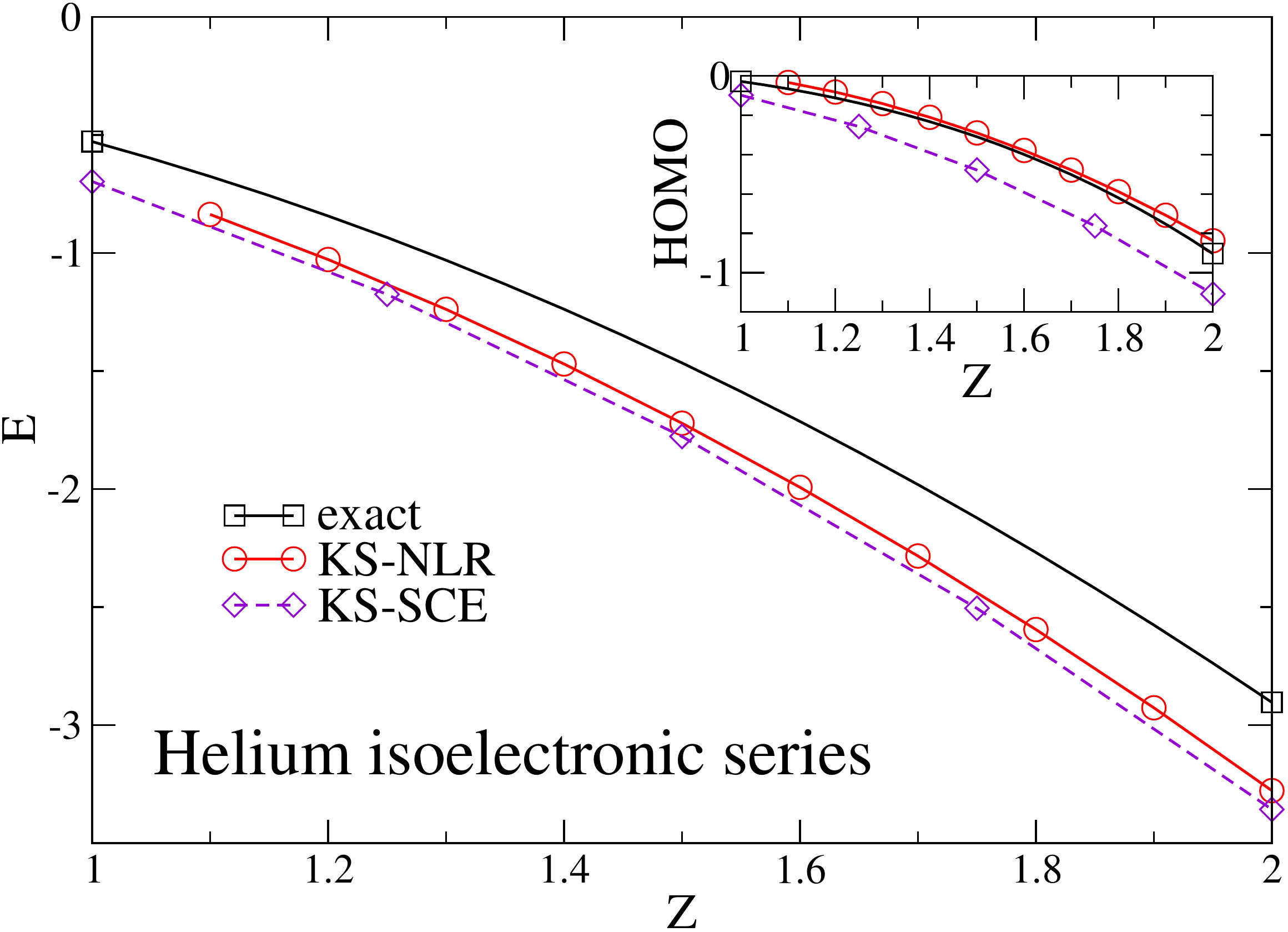}
\caption{%
Self-consistent energies and HOMO eigenvalues of the KS-NLR and KS-SCE functionals for
the 3d helium isoelectronic series ($N=2$, $Z$ variable), compared
with the exact behavior (cubic spline interpolation of $Z=1$, 2, 4, 10 values from \Ref{UG94}), 
KS-SCE data from \Ref{MirUmrMorGor-JCP-14}.
\label{3dZ2}
}
\end{figure}

\shd{Helium}
See \Figref{3dHe} for a self-consistent treatment of the helium atom, both with the
KS-NLR functional \eqref{EvKSNL} and the KS-SCE functional.
Both strong-correlation functionals gives a self-consistent energy
which is low compared to the exact (\mbox{$-$3.278} in KS-NLR and \mbox{$-$3.357} in KS-SCE \cite{MirUmrMorGor-JCP-14},
whereas the true energy of helium is \mbox{$-$2.904} \cite{UG94}), and both also give a helium density
much too contracted.  This contraction gives KS-SCE and KS-NLR
a larger kinetic energy
and a more-negative potential energy than the exact helium atom.  Comparing
the strong correlation methods:  the KS-SCE
and KS-NLR self-consistent densities are very similar;
the KS-NLR density decays a little more slowly due to a higher HOMO eigenvalue,
which can be seen in \Figref{3dZ2}.

\shd{Hydrogen anion}
While the NLR functional appears to be a lower bound for these finite systems,
this does not mean that the NLR functional always binds
when the true system does.  We work through this paradox by considering
the anion H$^-$.  
For the exact density $\n(\br)$ of the system,
the NLR energy is less than the true energy, $E_v\KSNL[\n] < E_v[\n]$,
since $W\xc\NL[\n] < E\xc[\n]$ in \Tabref{NLatoms}.  
Nevertheless, it appears that one can find a lower NLR energy $E_v\KSNL[\tilde \n]$ for some density
$\tilde \n(\br)$ which sends a fraction of an electron to infinity.  
With a large enough basis set, this can be deduced
from a positive HOMO eigenvalue in the self-consistent treatment, which we found when trying to 
converge H$^-$ for KS-NLR. 

Many standard DFT approximations behave exactly the same way for anions: in an infinite
basis set,
functionals such as B3LYP and PBE would send a fractional number of electrons to infinity \cite{KSB11}.  
Nevertheless, in \Ref{KSB11},
Burke and coworkers find that these functionals can yield good electron affinities, if the anion is calculated
with the functionals evaluated on densities with less error (in the atomic case Hartree--Fock densities).
The functional evaluated on a good anion density has a lower energy than the functional
self-consistently evaluated on the neutral system, giving a reasonable electron affinity
for standard functionals \cite{KSB11}.
As we have already seen, this is similar to the KS-NLR functional:
the non-self-consistent energy of H$^-$  is 
lower than the self-consistent KS-NLR energy of H, even though KS-NLR will not self-consistently
bind an extra electron.  With this approach, however, 
the KS-NLR electron affinity is a severe overestimation of the true electron affinity ($A\KSNL = 0.148$
whereas $A = 0.028$ \cite{UG94}, with $A = E_v(Z) - E_v(Z+1)$).

\begin{figure}
\includegraphics[width=\columnwidth]{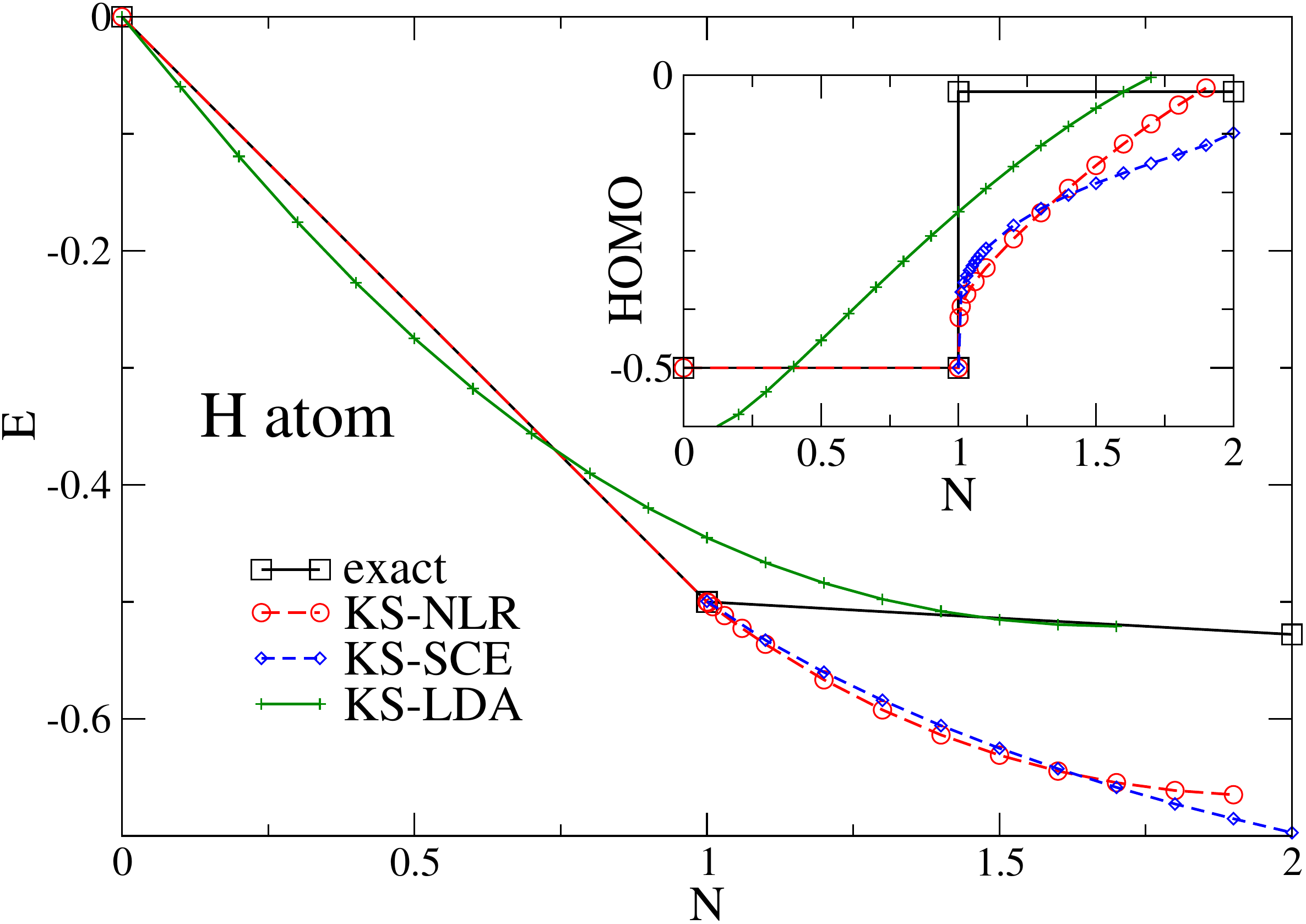}
\caption{%
Self-consistent energies of the KS-NLR, KS-SCE, and KS-LDA methods for the hydrogen atom
with a variable number of electrons $N$, performed within a spin-restricted framework.
KS-SCE and KS-LDA data from \Ref{MirUmrMorGor-JCP-14}.
}
\label{Z1PPLB}
\end{figure}

The KS-SCE functional requires no special treatment for anions as it is able to bind them, though as usual its energy
is far too low (see \Figref{Z1PPLB}), resulting in severe overbinding \cite{MirUmrMorGor-JCP-14}.  The KS-SCE HOMO, however is often a good approximation to the affinity \cite{MalMirGieWagGor-PCCP-14,MirUmrMorGor-JCP-14}. 

\shd{Variable $N$ with $Z=1$}
We can see how close KS-NLR comes to binding two electrons by considering a hydrogen atom
with a fractional number of electrons $N$. The exact energy as a function of $N$
should be piecewise linear with kinks at integer $N$  \cite{PPLB82}, and the HOMO
energy (which is the derivative of the energy with respect to $N$) should be a series of steps jumping at integer $N$.  
For hydrogen, the exact behavior of the energy and the HOMO eigenvalue is plotted in \Figref{Z1PPLB},
alongside KS-NLR, KS-LDA, and KS-SCE functionals.  When spin-restricted, KS-LDA makes a fractional spin
error for the neutral hydrogen atom ($N=1$, $N\up = N\dn = 1/2$) \cite{CMY08}, which has consequences
for dissociating H$_2$ -- we discuss this more for the 1d case later.  On the other hand,
KS-NLR and KS-SCE are exact for $N \le 1$, but err substantially for $N > 1$.  Nevertheless,
there is a nonanalyticity in the energy 
(and thus in the HOMO) energies at $N=1$ for KS-NLR and KS-SCE, unlike in KS-LDA, which is the sign that 
{some of the right} physics is being captured. 
{Within time-dependent DFT, this behavior is crucial}: 
non-analytic behavior at the integers is the key to to describe important phenomena as, e.g., charge transfer \cite{HelGro-PRA-12}.
The strong correlation functionals 
KS-NLR and KS-SCE bind for larger values of $N$ than KS-LDA, though KS-NLR stops short
of binding for $N=2$, as we have already noticed.  But as can be seen, KS-NLR energetically looks a lot like
KS-SCE, though intriguingly drops below KS-SCE for certain values of $N$.

\shd{Critical $Z$ with $N=2$}
To investigate further,
we decrease the nuclear charge $Z$ in fractional amounts from 2 to 1, moving from He to H$^-$
\cite{MirUmrMorGor-JCP-14}.  If there were ever such a thing as fractionally charged nuclei,
there is a critical value of $Z$ below
which two electrons would not bind:  $Z_c \approx 0.9110$ \cite{BFH90,EBMD14}.
In KS-NLR, however, we have already seen that the critical $Z$ value is above 1, since
hydrogen does not bind two electrons.
In \Figref{3dZ2} we plot our self-consistent results for fractional $Z$.  
Using our numerical approach, we could easily converge down to $Z=1.1$,
so we anticipate the critical $Z$ in KS-NLR near or just below that.
In KS-SCE, there is no difficulty in binding extra electrons, and $Z_c$ is severely underestimated:
$Z_c \approx 0.7307$ \cite{MirUmrMorGor-JCP-14}.

We remark here that, as usual, KS-NLR energies track quite well 
along with the KS-SCE energies, though there is a fairly large gap in the HOMO
energies due to the KS-NLR potential being higher than the KS-SCE potential.
Interestingly, this makes the KS-NLR HOMO eigenvalues fairly close to the exact  HOMO
eigenvalues, and suggests that electron ionization energies might be well
described by the KS-NLR HOMO energies.  We investigate this in other systems next.

\shd{Many-electron 3d atoms}
In \Tabref{NLatoms3d} we report total energies, ionization energies $I = E_v(N-1) - E_v(N)$, 
and HOMO
eigenvalues for atoms self-consistently calculated using the KS-NLR method.  
In exact KS-DFT, $\epsilon\homo$ should be equal to minus the exact, many body, ionization potential $I$ \cite{LevPerSah-PRA-84}.  
As expected from Fig.~\ref{Z1PPLB}, the KS-NLR method does not give good
agreement between its own ionization energies (computed by taking energy differences) and $-\epsilon\homo$. Instead, the KS-NLR HOMO energies 
are reasonably close to the true $I$ for the atoms in \Tabref{NLatoms}:  $-7$\%, $-19$\%, $-15$\%, and $-2$\% errors for He, Li, Be, and Ne,
respectively.  

\begin{table}
\begin{tabular*}{\columnwidth}{@{\extracolsep{\fill}}llccc}
\hline
atom & $E\KSNL$ & $I\KSNL$ & $-\epsilon\homo\KSNL$ & accurate $I$\\
\hline
He   & $-$3.278  & 1.278  &  0.84  &  0.904 \\
Li   & $-$8.170  & 0.279  &  0.16  &  0.198 \\
Be   & $-$15.76  & 0.45  &  0.29  & 0.343 \\
Ne   & $-$134.9  & 1.2  &  0.78 &  0.792 \\
\hline

\end{tabular*}
\caption{
Self-consistent atoms and ions within the KS-NLR method in 3d.
Exact ionization energies are experimental values from \Ref{CCCBDB}.
}
\label{NLatoms3d}
\end{table}

As the atomic number $Z$ gets larger, we expect KS-NLR to become asymptotically correct
for the ionization energy, since exchange and correlation energies show up at
smaller orders of $Z$ than kinetic, Hartree, and potential energies 
for large atoms \cite{LCPB09}.  The KS-NLR interaction XC energy scales like exchange,
though with a larger (in magnitude) coefficient.  To quantify this,
for large $Z$ atoms,
exchange dominates over correlation, and is locally 
$\epsilon\x(r\s) \approx -0.458/r\s$ \cite{D30},
about 1.6 times smaller than the uniform gas limit of KS-NLR \eqref{wxcNLunif}.
Non-uniform effects certainly play a role, but will not affect higher orders
of $Z$.

We now turn our attention to various 1d systems.

\ssec{Parabolic traps in 1d}
Of interest in strong correlation physics is the confinement of electrons 
in low-dimensional nanostructures such as quantum wires and quantum dots \cite{MalMirCreReiGor-PRB-13,MenMalGor-PRB-14}.
We will consider the parabolic trap of \Ref{MalMirCreReiGor-PRB-13} as a model for the quantum
wire, though with the soft-Coulomb interactions of \Ref{WSBW12}.  The
external potential is then given by $v(x) = k x^2 / 2$.
In a weakly correlated trap ($k \gtrsim 10^{-1}$ for our interactions),
the quantum kinetic energy operator dominates the physics of the electrons.
This regime yields Friedel-type oscillations of wavelength
$2 k_F$, where $k_F = \pi \bar \n / 2$ is the effective Fermi wavenumber,
with $\bar \n$ the effective average density in the middle of the trap \cite{MalMirCreReiGor-PRB-13}.
But as the confinement in the parabolic trap weakens ($k \ll 10^{-1}$),
the Coulomb repulsion operator drives the electron physics,
and we observe a $2 k_F \to 4 k_F$ transition in the wavelength of the
density oscillations.  Peaks form in the density where charge localizes,
and these peaks are the tell-tale signs of a Wigner-like regime. The challenge in this kind of systems is to capture this crossover without introducing magnetic order (i.e., without symmetry breaking), something that has been tried with GGA and self-interaction corrections without success \cite{VieCap-JCTC-10,Vie-PRB-12}.

\begin{figure}
\includegraphics[width=\columnwidth]{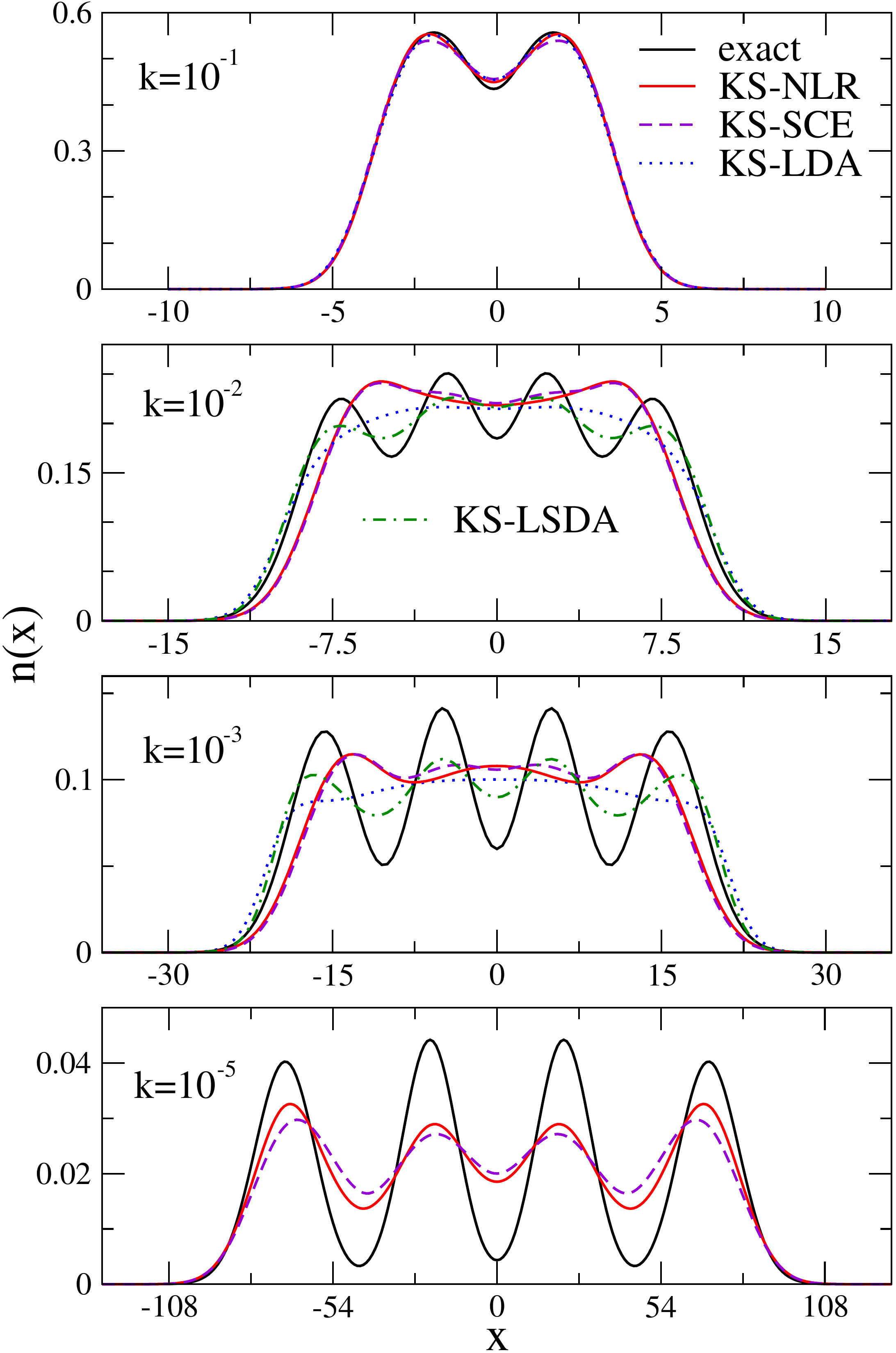}
\caption{%
Densities of the $2k_F \to 4k_F$ transition for $N=4$ soft-Coulomb interacting electrons in a parabolic 
external potential $v(x) = k x^2 / 2$.  
Exact density from
DMRG \cite{SWWB12}.
}
\label{1dHO}
\end{figure}

\begin{figure}
\includegraphics[width=\columnwidth]{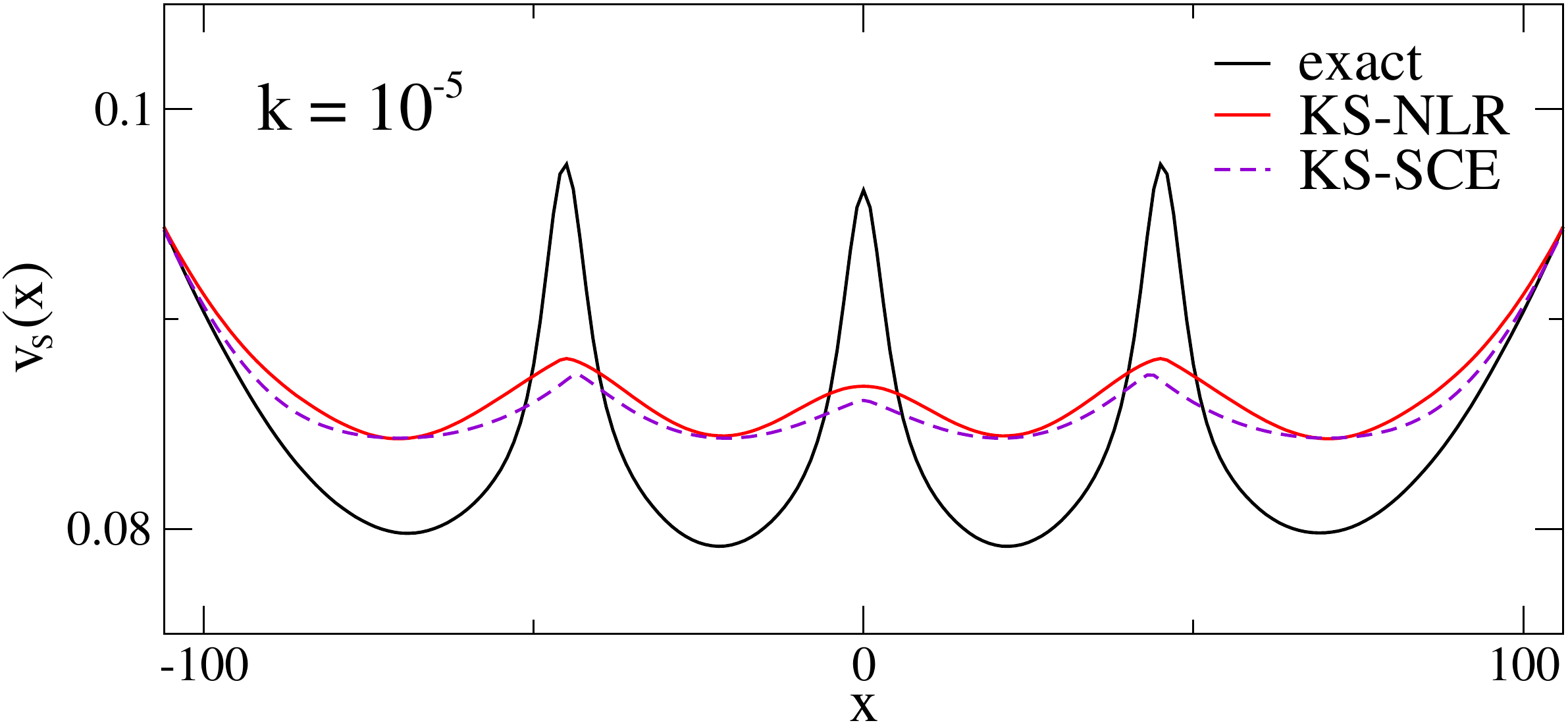}
\caption{%
KS potentials for the $k=10^{-5}$ system of \Figref{1dHO}.
}
\label{1dHOks}
\end{figure}

The KS-SCE functional captures this $2 k_F \to 4 k_F$ crossover \cite{MalMirCreReiGor-PRB-13}, and so does the KS-NLR functional, without any symmetry breaking.
We can see this in \Figref{1dHO}, which plots the densities of $N=4$ electrons
in parabolic traps with varying confinement strengths.  
As $k$ becomes small, the strongly correlated KS-SCE and KS-NLR methods naturally
produce peaks in the right locations.  Neither method develops the peaks as strongly
or as quickly as the exact result, but the KS-NLR method has a slight edge for very low densities.
Unfortunately, the KS-NLR functional predicts an
unphysical transition
region with 3 peaks near $k = 10^{-3}$, whereas the KS-SCE method correctly predicts
only either two or four peaks.  These density peaks are a result of barriers in the 
KS potentials \cite{Pb85}, which we plot in \Figref{1dHOks} for $k=10^{-5}$, and are well known to be of non-local nature \cite{HTR09}.

For comparison in \Figref{1dHO},
we also have 1d LDA results
using the correlation fit from \Ref{HFCV11} and exchange
from \Ref{WSBW12}.  While KS-LDA does well for weakly correlated
systems (large $k$), things get worse for small $k$.
Around $k=0.01$, KS-LDA looks like the Thomas--Fermi solution,
since the potential is very slowly varying.   But as $k$ gets even smaller,
KS-LDA becomes very difficult to converge, as the density becomes very delocalized and a very large grid in needed \cite{MalMirCreReiGor-PRB-13}.
By breaking spin-symmetry \cite{WSBW12}, the local {\em spin}-density
approximation (LSDA) method can achieve peaks \cite{APXT07}.
However, {these systems} have no magnetic order, so density functionals
should capture the physics  without breaking symmetry.
This is required on a more fundamental level for transition metal oxides above the N\'eel temperature, 
where magnetic order is destroyed but strong charge localization (the insulating phase) is still present 
\cite{AniZaaAnd-PRB-91}.
Notice that we could not converge either
KS-LDA or KS-LSDA for very small $k$, i.e.\ $\le 10^{-5}$. 
To date, 
only strongly correlated functionals like KS-SCE and KS-NLR correctly
localize charge -- {without introducing magnetic order} -- in regions
where the external potential offers no hints.

\ssec{Simple 1d molecules}
Here we investigate the binding energy curves of various 1d molecules, where
we consider the total energies of the system for soft-Coulomb interacting systems.
Thus we add the electronic energy and the interaction energy between nuclei:
\ben
E_0(R) \equiv E_v + \dfrac{Z_1 Z_2}{\sqrt{R^2+1}},
\een
where $v(x) = -Z_1 / \sqrt{x^2+1} - Z_2 / \sqrt{(x-R)^2+1}$.  This soft 1d universe
is a laboratory to test functionals and ideas about correlation \cite{WSBW12}, where
we have easy access to exact answers using the density matrix renormalization group \cite{SWWB12}.  We will consider neutral systems with $N = Z_1 + Z_2$.

\shd{1d hydrogen molecule}
We consider the most infamous example of bond breaking in KS-DFT:
H$_2$.
Standard KS-DFT methods fail to dissociate H$_2$ correctly because
of {\em fractional spin error}:
the energy of a single hydrogen atom with one spin-up (or one spin-down) electron is different
than with half an up-spin  and half a down-spin electron \cite{CMY08}.
At dissociation, H$_2$ comprises two such spin-unpolarized atoms,
whereas functionals typically give accurate values for a single H atom 
only when spin-polarized.
This difficulty occurs for all molecules which dissociate into open-shell fragments.

\begin{figure}
\includegraphics[width=\columnwidth]{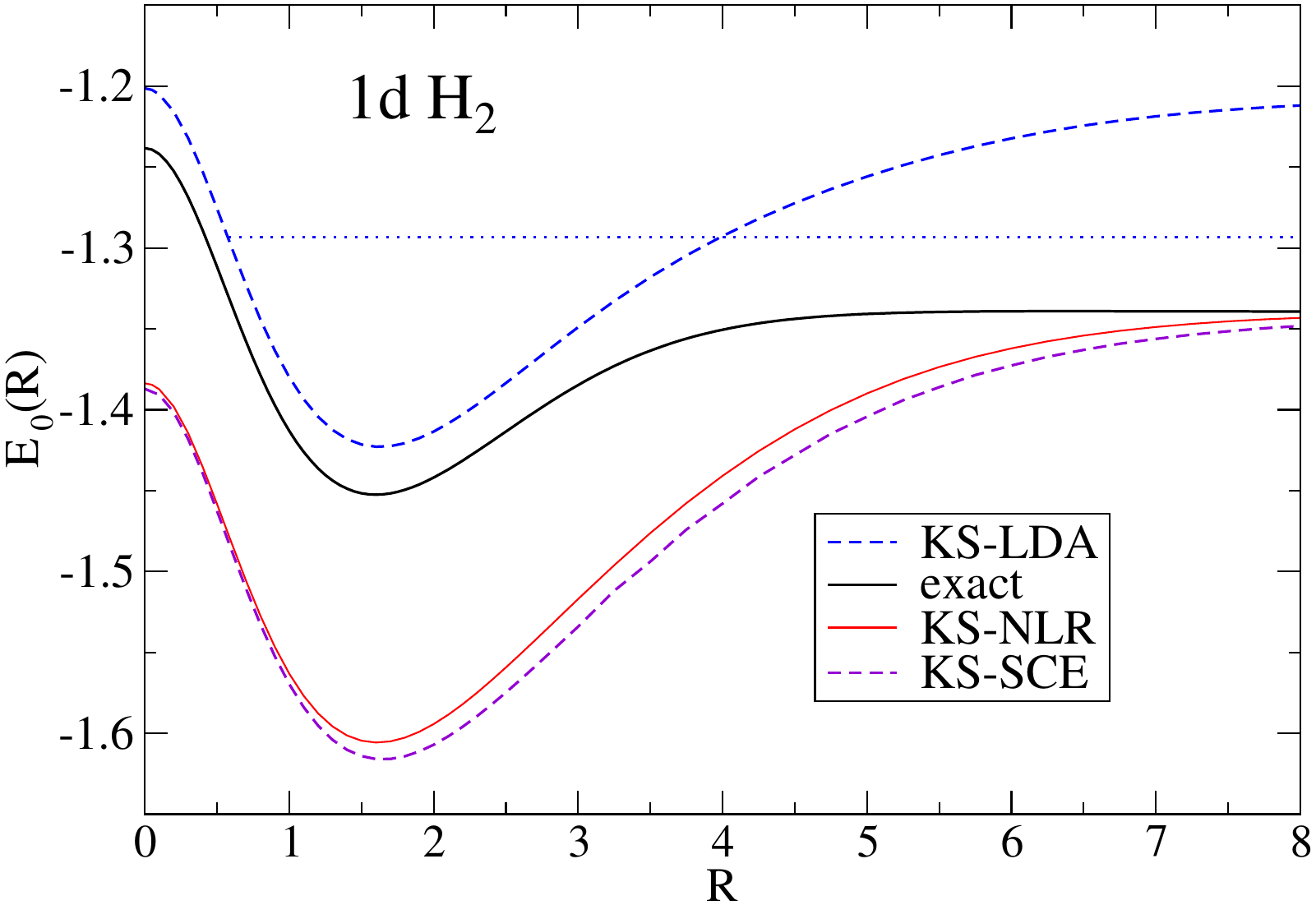}
\caption{%
The total molecular energy of 1d H$_2$ with soft-Coulomb interactions between all particles.
Exact and KS-LDA data from \Ref{WSBW12}.  The horizontal dotted line is twice the energy
of a KS-LSDA hydrogen atom.
}
\label{1dH2}
\end{figure}

In \Figref{1dH2} we plot the binding energy curve of H$_2$ for various functionals.
At dissociation, KS-LDA errs due to its fractional spin error, whereas
both strongly correlated methods
KS-SCE and KS-NLR dissociate H$_2$ correctly, i.e.\ $E_0(R) \to 2 E_{\rm H}$ as $R\to\infty$.
As in the parabolic 1d traps,
breaking spin symmetry allows KS-LSDA to dissociate correctly \cite{WSBW12,CF49} (not shown in \Figref{1dH2}, 
but see \Refs{WSBW12,MalMirGieWagGor-PCCP-14}), 
but with all the caveats mentioned for breaking spin symmetry in the parabolic traps. 
However, both strong correlation functionals bind the H$_2$ molecule much too strongly, and the
well extends out to too large of $R$.  Despite these gross chemical inaccuracies, the equilibrium bond length is
overestimated by only 1\% in KS-NLR and 3\% in KS-SCE, whereas
KS-LDA makes a 2\% error \cite{WSBW12}.  

We emphasize here that both KS-SCE and KS-NLR do not need to break
spin symmetry to dissociate H$_2$ correctly.  In fact, there is no spin dependence in either
the NLR or SCE functional, so neither functional can lower the energy by breaking spin symmetry.
Instead, breaking spin symmetry {\em raises} the kinetic energy of the KS wavefunction, and there is
no corresponding decrease in the interaction energy (since it is spin-independent).  Thus a spin-unrestricted
calculation within either KS-NLR or KS-SCE yields the same result as a spin-restricted calculation.  See also
\Ref{MenMalGor-PRB-14} for a similar discussion with KS-SCE applied to parabolic traps.

\begin{figure}
\includegraphics[width=\columnwidth]{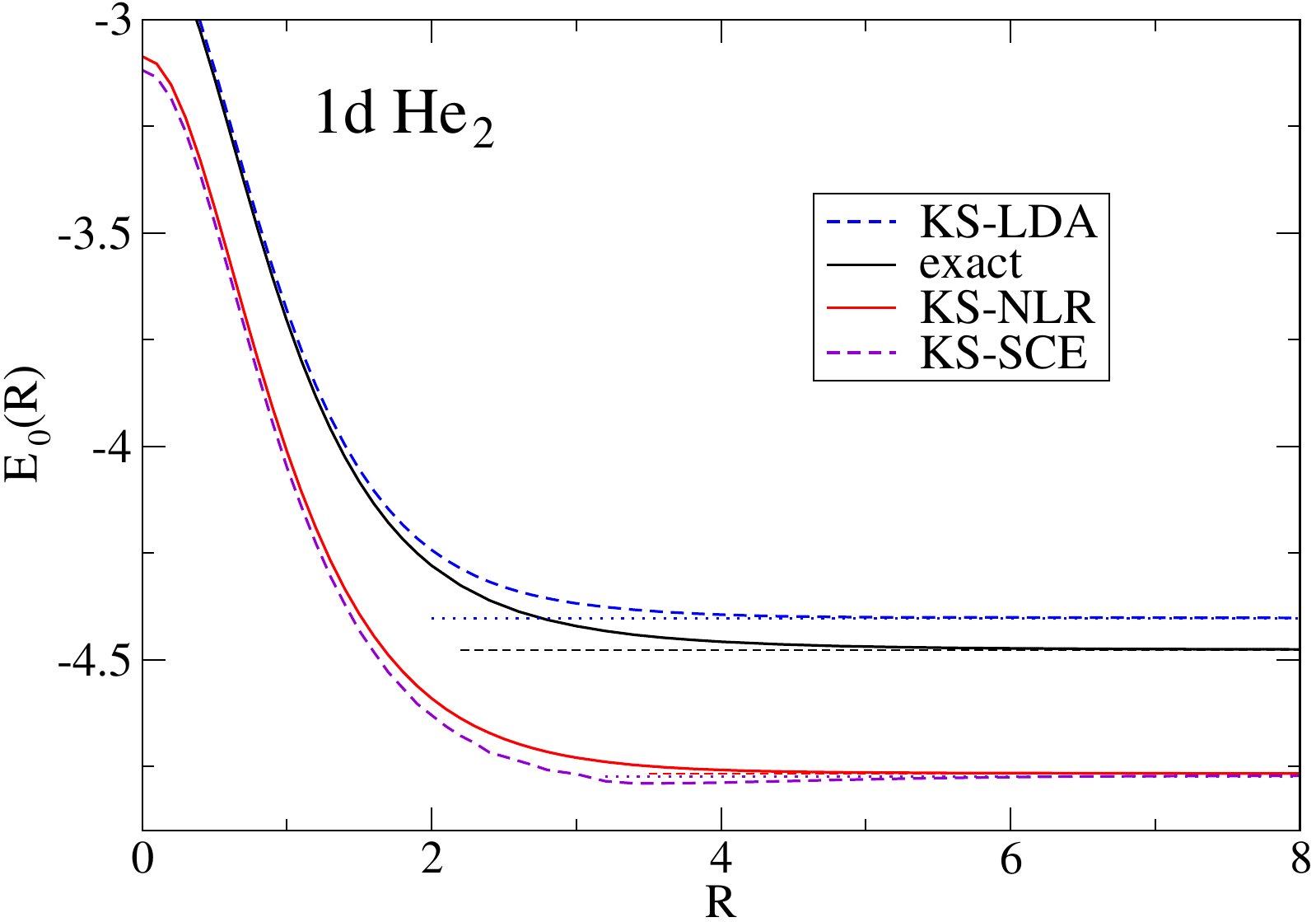}
\caption{%
(Lack of) binding for the 1d soft-Coulomb helium dimer.  KS-SCE predicts a weakly bound state.
Horizontal lines indicate twice the energy of a single isolated helium atom within each method.
}
\label{1dHe2}
\end{figure}

\shd{1d helium dimer}
We now consider a molecule which dissociates into closed-shell fragments, He$_2$.
Here the issues of fractional spin disappear, since the fragments are closed-shell.
Dissociating He$_2$ yields two spin-unpolarized helium atoms, and standard KS-DFT approximations do well
for spin-unpolarized helium atoms.  See \Figref{1dHe2} for plots of the 1d He$_2$ binding energy.
KS-LDA dissociates correctly (to the value of $2 E_{\rm He}\KSLDA$), as do the strongly correlated methods.
In 3d, there is a very weak van der Waals bond for the helium dimer;
but soft-Coulomb 1d He$_2$ has no bound state.  Here KS-SCE incorrectly predicts a weakly bound state.

\begin{figure}
\includegraphics[width=\columnwidth]{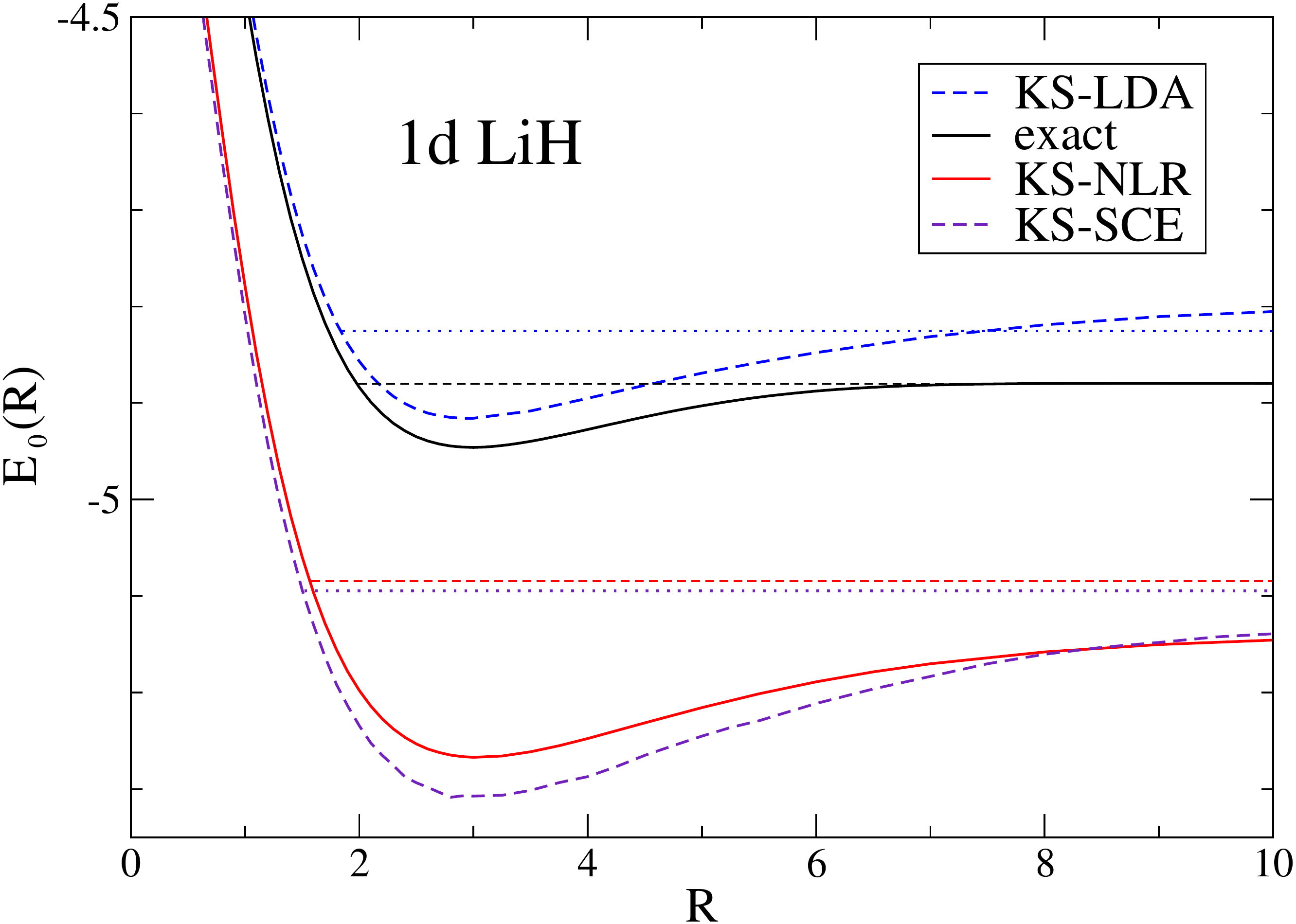}
\caption{%
The total molecular energy of 1d LiH with soft-Coulomb interactions between all particles.
Horizontal lines indicate the sum of energies of an isolated Li atom and an isolated H atom
within each method.  
}
\label{1dLiH}
\end{figure}

\shd{1d lithium hydride}
Here is another example of dissociating into open-shell fragments:  lithium ($Z_1 = 3$)
hydride ($Z_2 = 1$).  See \Figref{1dLiH} for a plot of the binding energy curve.
Much of the discussion on H$_2$ carries over for LiH.  KS-LDA
makes an error in the dissociation limit due to fractional-spin error,
whereas KS-SCE and KS-NLR do not.  Both KS-NLR and KS-SCE methods overbind.
Here we found it challenging to converge the KS-SCE method for large $R$ values,
and there is some numerical noise in the KS-SCE data due to the integration methods used.
As can be seen in \Figref{1dLiH}, KS-NLR is going very slowly to its
dissociation limit.  This is due to the long-ranged behavior of the
{NLR interaction XC energy per particle, $w\xc\NL(x)$, as seen in
e.g.\ \Figref{3dHewxc} (and see nearby discussion).}

\sec{Future work and conclusions}

This work lays the mathematical foundation for the application
of an approximate functional (called NLR) for the strong-interaction limit of DFT, using the nonlocal
Wigner--Seitz radius $R(\br)$.
This functional is capable of
dissociating single bonds and localizing charge density, key signatures
of strong correlation physics, and it is computationally much more accessible than the exact strong-correlation limit of DFT, SCE. The energy as a function of the number of electrons $N$ displays non-analyticities at integer $N$ even for open shell systems in a spin-restricted formalism, although we do not fully capture the correct energy versus $N$ piecewise linear behavior.
In its present form, however, the functional may not
bind negative ions, and its potential does not have the right asymptotic behavior.
We envision many avenues of future research, which build upon the ideas
presented in this work and {seek to correct the deficiencies of the NLR functional.}

The main challenge is to build corrections to our functional that add the accuracy of standard DFT for weakly and moderately correlated systems without destroying the ability of NLR to capture strong-correlation features, and, possibly, to improve them.
One can produce corrections to make the nonlocal method correct for the uniform gas \cite{MalMirCreReiGor-PRB-13},
as well as one-electron systems,
by using the nonlocal Wigner--Seitz radius $R(\br)$ in place of (or somehow
combined with) the local $r\s(\br)$
in the LDA XC energy density.  Since $R(\br) \to 0$ for one-electron systems,
the XC energy density will also go to zero.  Despite the simplicity of this approach,
there are many free parameters which affect the transition from one electron to many.
This nonlocal LDA is the subject of current work.
Suitable corrections may also fix the asymptotic behavior of the
NLR XC potential.

Using the energy density of the NLR functional to create a local interpolation along the adiabatic connection, using the exact (or an approximate) exchange energy density for the weakly correlated limit seems also a very promising route of action. 

The introduction
of spin densities usually helps in obtaining better energies and to treat open shell systems.  There are many
ways to generalize the nonlocal machinery to spin-polarized systems. 
Here we suggest just one way:  continue to define the nonlocal radius $R(\br)$
using the total density, but then define the polarization via an
average of the local polarization within the nonlocal Wigner--Seitz sphere $\Omega(\br)$:
\ben
\zeta(\br) = \intOrp \big( \n\up(\br') - \n\dn(\br') \big)
\een
This will equal 1 if $\n\up(\br) = \n(\br)$ in the Wigner--Seitz sphere $\Omega(\br)$,
will equal $-1$ if $\n\dn(\br) = \n(\br)$, 
and will otherwise land somewhere in between. For an unpolarized electron density,
$\n\up(\br) = \n\dn(\br)$ so $\zeta(\br) = 0$.  Using the nonlocal $\zeta(\br)$ in place of the
local version might well complement the nonlocal LDA strategy outlined above.

For a completely different application, one can use $R(\br)$ to develop a nonlocal orbital-free kinetic energy functional
(as in the original paper \cite{H86} which inspired the definition of $R(\br)$), by writing
the Thomas--Fermi kinetic energy using $R(\br)$ and/or combining with the
von Weizs\"acker kinetic energy.  Again there are many free parameters
even after constraints to yield the correct limits
for the uniform gas and one-electron systems.

\shd{Acknowledgments}
We thank Andr\'e Mirtschink, Andreas Savin, Stefan Vuckovic, Giovanni Borghi,
and Viktor Staroverov
for insightful discussions.
We gratefully acknowledge funding through an NWO Vidi grant,
and LOW appreciates additional support through DOE grant DE-SC0008696.

\appendix

\sec{Calculating $R(\br)$}\label{calculateR}

In general, to calculate $R(\br)$ we need to be able to integrate the density in a sphere
concentric with the point $\br$ with an arbitrary radius $R$:
\ben
N_e(\br,R) \equiv \intrp \n(\br')\, \heavi\big( R - |\br-\br'| \big).\label{NerR}
\een
If we can fit the density to a sum of Gaussian or Slater-type exponentials, this integral
becomes an analytical function of $R$.
An alternative, basis-independent approach is to use information from the Hartree potential.
We can then obtain $R(\br)$ by finding the root $N_e(\br,R(\br)) = 1$.
Since $N_e(\br,R)$ is monotonically increasing in $R$, 
we easily obtain the root by increasing the radius from zero or by using Newton's shooting method.

\shd{Evaluating $N_e(\br,R)$ for Gaussian-type densities}
Imagine that your density is well described by a single
Gaussian centered at $\hat{z} b$ with decay constant $\alpha$,
i.e.\ $n(\br) = e^{-\alpha |\br - \hat{z}b|^2}$.
(Later we will sum over many such terms.)
We now integrate the density in a sphere of radius
$R$ concentric with the origin.
The number of electrons in that
sphere is:
\ben
N^{G,\alpha}_b(R) \equiv 
                    \intr e^{-\alpha |\br - b\hat{z}|^2}  \heavi ( R - |\br| ). \label{NGdef}
\een
This may be evaluated quite easily:
\bea
N^{G,\alpha}_b(R) &=& -\pi \dfrac{1-e^{-4 \alpha b R}}{2 \alpha^2 b} e^{-G_-^2}  + \nonumber \\
&& \half \left( \dfrac{\pi}{\alpha}\right)^\frac{3}{2}\big( \erf G_+ + \erf G_- \big)
\eea
where $G_\pm \equiv \sqrt{\alpha}(R \pm b)$.
Alternatively, we can first calculate the density of electrons integrated over
the surface of a sphere of radius $R$, where the center of the sphere is displaced
a distance $b$ from the center of the Gaussian of decay $\alpha$:
\ben
S^{G,\alpha}_b(R) \equiv  
R^2 \intusph e^{-\alpha |R \hat u - b\hat{z}|^2} .
\een
This analytically integrates to:
\ben
S^{G,\alpha}_b(R)=
 \dfrac{\pi R}{\alpha b} e^{-G_-^2} \big(1 - e^{-4 \alpha b R} \big).
\een
Then we can calculate $N^{G,\alpha}_b(R)$ by integrating $S^{G,\alpha}_b(R')$
from $R' = 0$ to $R$.

We now use this information in order to calculate $R(\br)$ 
when the density is a sum of such Gaussians:
\ben
\n(\br) = \sum_j g_j e^{-\alpha_j |\br - \bR_j|^2 }. \label{gaussdens}
\een
For this density,
we now  find the number of electrons in a sphere
of radius $R$, centered at $\br$.  To do this, we use
\Eqref{NGdef} and shift the origin for each term in the sum
\eqref{gaussdens}:
\ben
N_e(\br,R) = \sum_j g_j\, N^{G,\alpha_j}_{|\br-\bR_j|} ( R ). \label{AddGaussians}
\een
And we find $R(\br)$ by finding the root $N_e(\br,R(\br))=1$ as discussed
earlier.

\shd{Evaluating $N_e(\br,R)$ for Slater-type densities}
Similarly, we evaluate the number of electrons in a sphere of radius $R$
at a distance $b$ away from a Slater-type function with decay constant $\alpha$:
\ben
N^{S,\alpha}_b(R) \equiv \intr e^{-\alpha |\br - b\hat{z}|}  \heavi( R - |\br| ).
\een
And by evaluation:
\ben
N^{S,\alpha}_b(R) = \left\{
\begin{array}{lll}
N^S_< & & R < b \\
N^S_> & & \text{otherwise} \\
\end{array}
\right.
\een
where
\bea
N^S_< &=& \dfrac{4 \pi e^{-\alpha b}}{\alpha^4 b} \Big( \alpha R\, (3 + \alpha b) \cosh \alpha R  + \nonumber \\ 
&&\quad \quad \quad \quad \quad -(3 + \alpha b + \alpha^2 R^2 ) \sinh \alpha R \Big),\quad
\eea
and
\bea
N^S_> &=& \dfrac{8\pi}{\alpha^3} + \dfrac{4 \pi e^{-\alpha R}}{\alpha^4 b} \Big( \alpha b (1 + \alpha R) \cosh \alpha b + \nonumber \\
&&\quad \quad\quad\quad\quad - (3 + 3 \alpha R + \alpha^2 R^2 ) \sinh \alpha b \Big).\quad
\eea

If we expand the density in a sum of Slater-type functions:
\ben
n(\br) = \sum_j s_j e^{-\alpha_j |\br-\bR_j|}, \label{slaterdens}
\een
then $N_e(\br,R)$ is a simple sum:
\ben
N_e(\br,R) = \sum_j s_j\, N^{S,\alpha_j} _{|\br - \bR_j|}  (R ). \label{AddSlaters}
\een

\shd{Evaluating $N_e(\br,R)$ using Gauss' law}
Gauss' law allows us to determine the charge $Q$ contained in a volume $\Omega$ by integrating
the electric field ${\bf E}(\br)$ permeating the surface $\partial \Omega$ of the volume.  In atomic units:
\ben
Q = \dfrac{1}{4 \pi} \intdOr {\bf E}(\br).
\een
For electrons, the field is the gradient
of the Hartree potential:  ${\bf E}(\br) = \nabla v\H[\n](\br)$.  
Taking care with signs, the number of electrons in a sphere
of radius $R$, centered at $\br$ is:
\ben
N_e(\br,R) = -\dfrac{R^2}{4\pi} \intusph \dfrac{\partial v\H[\n](\br + R \hat u)}{\partial R}.
\een
We have not checked the efficiency of this approach, but we include it here 
in case it proves useful.

\sec{Evaluating functional derivatives}\label{calculatev}

Here we derive $v\xc\NL[\n](\br) = \delta W\xc\NL[\n] / \delta \n(\br)$ 
by first determining how the NL radius $R(\br)$ changes when the density changes
by a small amount $\delta \n(\br)$.
One may implicitly differentiate the definition of $R(\br)$ in \Eqref{Rdef} and obtain:
\ben
\delta R(\br) =  \dfrac{-1}{S(\br)}\intOrp \delta n(\br')  , \label{deltaR}
\een
where $\Omega(\br)$ is the sphere defined by the original Wigner--Seitz radius $R(\br)$,
and  $S(\br)$ is the {\em nonlocal radial density} (with units of length$^{-1}$), 
defined by integrating
the density over the surface $\partial \Omega(\br)$ of the nonlocal Wigner--Seitz sphere, which can be performed in many diferent ways:
\bea
S(\br) &\equiv& \intSrp  \n(\br') \label{S} \\
&=& R^2(\br)\intusph \n(\br + R(\br) \hat u) \\
&=& \intrp \n(\br')\, \delta\big( R(\br) - |\br - \br'| \big)\label{SDirac}  \\
&=& \left.\dfrac{\partial N_e(\br,R)}{\partial R}\right |_{R = R(\br)}.
\eea
To use the chain rule in functional derivatives,
we rewrite \Eqref{deltaR} as:
\ben
\dfrac{\delta R(\br)}{\delta \n(\br')} =  \dfrac{-1}{S(\br)} \heavi\big(R(\br) - |\br - \br'|\big).
\label{deltaRdeltan}
\een

Now to determine the NLR XC potential.  The density appears in many different places inside
$W\xc\NL[\n]$, so we get a few different terms as the density varies:
\bea
&& v\xc\NL[\n](\br)  \\ 
&& = -\half \dfrac{\delta}{\delta \n(\br)} \int d^3 r' \int d^3 r'' \, 
                                                    \dfrac{\n(\br') \, \n(\br'')}{|\br' - \br''|}
                                                    \, \heavi\big(R(\br') - | \br' - \br'' | \big) \nonumber \\
&&= -\half \int d^3 r'' \, \dfrac{ \n(\br'')}{|\br - \br''|} \,
                            \heavi\big(R(\br) - | \br - \br'' | \big) \nonumber \\  
&&\quad- \half \int d^3 r' \, \dfrac{ \n(\br')}{|\br' - \br|} \,
                            \heavi\big(R(\br') - | \br' - \br | \big) \nonumber \\
&&\quad - \half \int d^3 r'\, \dfrac{\delta R(\br')}{\delta \n(\br)} \int d^3 r'' \, 
                        \dfrac{\n(\br') \, \n(\br'')}{|\br' - \br''|} \,
                        \delta\big(R(\br') - | \br' - \br'' | \big) . \nonumber
\eea
The first two terms can be combined using $g\xc\NL(\br,\br')$, and the third term we use our
new relation from \Eqref{deltaRdeltan} 
as well as the $\delta$ function to collapse $|\br' - \br''| \to R(\br')$:
\bea
&& v\xc\NL[\n](\br)  \\ 
&& = \int d^3 r' \, \dfrac{ \n(\br')}{|\br - \br'|}\, g\xc\NL(\br,\br') \nonumber \\
&& \quad - \half \int d^3 r'\, \left[ \left(  \dfrac{-1}{S(\br')} \heavi \big(R(\br') - |\br - \br'| \big) \right)
\dfrac{\n(\br') }{R(\br')} \right]\times \nonumber \\
&&\quad\quad\quad\quad\quad \int d^3 r'' \, \n(\br'')   \, \delta\big(R(\br') - | \br' - \br'' | \big)   \nonumber  \\
&&= \int d^3 r' \, \dfrac{ \n(\br')}{|\br - \br'|}\, g\xc\NL(\br,\br') \nonumber \\
&& \quad+ \half \int d^3 r'\, \left[  \dfrac{\n(\br')}{S(\br')\, R(\br')}  \,\heavi \big(R(\br') - |\br - \br'| \big) \right] \,  S(\br'), \nonumber 
\eea
where to get to the last line we used \Eqref{SDirac}.
Thus the $S(\br')$ cancels in this second integral, leaving us with \Eqref{vxcNL}.

\bibliography{../lmaster,../biblioPaola}

\end{document}